\documentclass[%
reprint, %
superscriptaddress,
showpacs,
amsmath,amssymb, %
aps, %
prd, %
twocolumn %
]{revtex4-1}
\usepackage{hhline}
\usepackage{xcolor}
\usepackage{amsmath}
\usepackage{hyperref}
\usepackage{graphicx}
\usepackage{bm}
\usepackage{dcolumn}
\begin{document}


\title{Wormhole geometries supported by three-form fields}

\author{Bruno J. Barros}
\email{fc36666@alunos.fc.ul.pt}
\affiliation{Instituto de Astrof\'isica e Ci\^encias do Espa\c{c}o,\\ 
	Faculdade de Ci\^encias da Universidade de Lisboa,  \\ Campo Grande, PT1749-016 
	Lisboa, Portugal}

\author{Francisco S. N. Lobo}
\email{fslobo@fc.ul.pt}
\affiliation{Instituto de Astrof\'isica e Ci\^encias do Espa\c{c}o,\\ 
	Faculdade de Ci\^encias da Universidade de Lisboa,  \\ Campo Grande, PT1749-016 
	Lisboa, Portugal}
\affiliation{Departamento de F\'{\i}sica, Faculdade de Ci\^{e}ncias da Universidade de Lisboa, Edif\'{\i}cio C8, Campo Grande, P-1749-016 Lisboa, Portugal}

\date{\today}

\begin{abstract}
In this work, we find novel static and spherically symmetric wormhole geometries using a three-form field. By solving the gravitational field equations, we find a variety of analytical and numerical solutions and show that it is possible for the matter fields threading the wormhole to satisfy the null and weak energy conditions throughout the spacetime, when the three-form field is present. In these cases, the form field is responsible for supporting the wormhole and all the exoticity is confined to it. Thus, the three-form curvature terms, which may be interpreted as a gravitational fluid, sustain these wormhole geometries. We also show that in the case of a vanishing redshift function the field can display a cosmological constant behavior.
\end{abstract}

\maketitle


\section{Introduction}\label{introduction}

Wormholes \cite{Morris:1988cz, Visser:1995cc} are tunnels connecting different regions of spacetime and have been a subject of discussion for almost a century, ranging from the Flamm solution \cite{Flamm}, the Einstein-Rose bridge \cite{PhysRev.48.73}, the geon concept devised by Wheeler \cite{Wheeler:1955zz,Misner:1957mt}, the Ellis \cite{Ellis:1973yv} and Bronnikov \cite{Bronnikov:1973fh} solutions in the 1970s, to the Morris-Thorne renaissance in 1988 \cite{Morris:1988cz}. 
A fundamental feature of wormhole physics is the flaring-out condition of the throat, which in General Relativity (GR), and through the Einstein field equations, entails the violation of the null energy condition (NEC) at the vicinity of the throat \cite{Morris:1988cz,Visser:1995cc,Lobo:2004rp,Curiel:2014zba,Lobo:2017oab,Barcelo:2000zf}. This leads to the assumption that, in order to sustain a wormhole, an exotic type of fluid, defined as matter that violates the NEC, must be at play. However, one may tackle the problem in the context of modifications of GR.
Indeed, it has been shown that it is possible to support wormhole geometries in modified gravity, where the matter threading the wormhole satisfies all of the energy conditions, and it is the effective energy-momentum tensor, containing higher order curvature derivatives, that is responsible for the NEC violation. Thus, the higher order curvature terms, interpreted as a gravitational fluid, sustain these wormhole geometries \cite{Harko:2013yb}. More specifically, a number of works have been dedicated to finding specific solutions in modified gravity, namely, in $f(R)$ gravity \cite{Lobo:2009ip}, Weyl gravity \cite{Lobo:2008zu}, curvature-matter couplings \cite{Garcia:2010xb,MontelongoGarcia:2010xd}, modified teleparallel gravity \cite{Bohmer:2011si}, Einstein-Gauss-Bonnet gravity \cite{Mehdizadeh:2015jra} and the hybrid metric-Palatini theory \cite{Capozziello:2012hr}, amongst many other scenarios.

In a cosmological context, an alternative to modified gravity as a possible cause of the recent accelerated expansion of the Universe \cite{Perlmutter:1998np,Riess:1998cb}, the inflationary phase of the early universe \cite{Guth:1980zm,Linde:1983gd,Martin:2013tda} and many other cosmological phenomena, is {\it dark energy} \cite{Wetterich:1994bg,Zlatev:1998tr,Barros:2018efl}, which can be formally represented by a scalar field. In the context of compact objects, such as in wormhole geometries, it is also possible that scalar fields are one of the simplest candidates to perform the role of exotic matter \cite{Barcelo:2000zf,Butcher:2015sea}. In fact, an extensive amount of work has been dedicated to wormhole physics supported by scalar fields, especially in the context of the stability issue. However, in wormhole geometries, the non-trivial topology is supported by a phantom scalar field, in order to satisfy the flaring-out condition. In particular, rotating wormhole solutions in GR were presented, which are supported by a phantom scalar field \cite{Kleihaus:2014dla,Kleihaus:2017kai}. These specific solutions evolved from a static Ellis wormhole configuration, when the throat is set into rotation, and as the rotational velocity increases, the throat deforms until an extremal Kerr solution is obtained, at a maximal value of the rotational velocity. Since the stability analysis of rotating wormholes in four dimensions is very involved, a stability analysis of five-dimensional rotating wormholes was performed with equal magnitude angular momenta only, by restricting the analysis to the unstable radial modes. Interestingly, when the rotation is sufficiently fast, the radial instability disappears for these five-dimensional wormholes.

In this work, and motivated by the analysis in scalar fields, we are essentially interested in finding wormhole geometries supported by three-forms, where the matter threading the wormhole satisfies the energy conditions. We emphasize that three-form fields \cite{Wongjun:2017spo,Koivisto:2009ew} are widely used in the literature and seem to present viable solutions to cosmological scenarios such as the recent acceleration of the Universe \cite{Morais:2016bev,Koivisto:2012xm,Koivisto:2009fb} and inflation \cite{Koivisto:2009sd,Kumar:2016tdn,DeFelice:2012jt,DeFelice:2012wy,Kumar:2014oka}. 
In fact, the cosmology of self-interacting three-forms was investigated, and it was shown that the minimally coupled canonical theory can naturally generate a variety of isotropic background dynamics, including scaling, possibly transient acceleration and phantom crossing \cite{Wongjun:2017spo,Koivisto:2009ew}.
In \cite{Koivisto:2009fb}, the background dynamics and linear perturbations of self-interacting three-form cosmology was investigated. It was shown that the phase space of cosmological solutions possesses (super)-inflating attractors and saddle points which can describe three-form driven inflation or dark energy. In \cite{Barros:2015evi}, a model of inflation was studied on a 5D Universe driven by a single three-form field confined to a 4D brane. The dynamics was studied by testing the perturbation variables, tensor to scalar ratio and spectral index, against observational data. Screening mechanisms have also been explored in the context of three-forms \cite{Barreiro:2016aln}. Considering a three-form field with conformal couplings to the matter sector, it was shown that it is possible to obtain a thin-shell setting where the field interactions are short range. In \cite{Turok:1998he}, Hawking and Turok showed that the inclusion of a 4-form, constructed from a three-form field potential, can naturally explain the vanishing cosmological constant problem.

Thus, in this work we analyze the aftermath of theoretically constructing a three-form field tailored for a static and spherically symmetric wormhole geometry. The aim of the present work is to find analytical and numerical solutions in which the matter fields satisfy the energy conditions throughout the entire wormhole spacetime, and it is the three-form field that is responsible for sustaining the wormhole, consequently violating all the energy conditions.

This paper is outlined in the following manner: In Section \ref{model} we present the model, write the action, introduce the three-form formalism, present the equations of motion for the field and finally compute the gravitational field equations. In Section \ref{ec}, we present an analysis regarding the energy conditions, in the presence of a three-form field. In Section \ref{solutions}, we expound the modus operandi for finding solutions, and explore their significance with analytical and numerical methods. Finally, we conclude in Section \ref{conclusions}.

\section{Wormhole geometries supported by three-forms}

\subsection{The metric and action}

\label{model}
We consider a static and spherically symmetric wormhole configuration, described by the following line element \cite{Morris:1988cz}
\begin{equation}
\label{metric}
ds^2 = -e^{2\Phi(r)} dt^2 + \frac{dr^2}{1-b(r)/r} + r^2 \left( d \theta^2 + \sin^2 \theta d\phi^2 \right),
\end{equation}
where $\Phi(r)$ is known as the redshift function, as it is related to the gravitational redshift, and is assumed to be finite everywhere in order to avoid the presence of event horizons, thus rendering the wormhole traversable \cite{Butcher:2015sea,Lobo:2009ip}. $b(r)$ is denoted the shape function, as it depicts the form of the wormhole. The radial coordinate $r$ runs from a minimum value $r_0$, corresponding to the throat of the wormhole, where $b(r_0)=r_0$, to $+\infty$. The divergence of the $g_{rr}$ component on the metric, Eq.~\eqref{metric}, triggers a coordinate singularity, so that we require the proper radial distance, $l(r) = \pm \int_{r_0}^{r} [1-b(r)/r]^{-1/2}dr$,
to be finite everywhere \cite{Morris:1988cz}. Note that the additional condition $b(r)\leq r$ is also imposed. 

Now, a key ingredient of wormholes is the so-called flaring-out condition \cite{Morris:1988cz}, given by $b'(r)<b(r)/r$, at the vicinity of the throat, where a prime denotes a derivative with respect to the radial coordinate $r$ \cite{Morris:1988cz}. This constraint entails information on the shape of the wormhole, expressed as constrains on $b(r)$, and reduces to $b'(r_0)<1$ at the throat.

The action of our model is described by
\begin{equation}
\label{action}
\mathcal{S}=\int d^4x\sqrt{-g}\left( \frac{1}{2\kappa^2}R + \mathcal{L}_{A} \right) + \mathcal{S}_{m}\left( g_{\mu\nu}, \psi \right),
\end{equation}
where $g$ is the determinant of the metric, i.e., $g\equiv{\rm det }\,\,g_{\mu\nu}$, $\kappa^2\equiv 8\pi G$, $R$ is the Ricci scalar and $\mathcal{S}_{m}\left( g_{\mu\nu}, \psi \right)$ is the matter action, where $\psi$ collectively defines the matter fields. We have introduced the existence of a three-form field $A_{\alpha\beta\gamma}$ \cite{Wongjun:2017spo,Koivisto:2009ew}, with the following Lagrangian density
\begin{equation}
\label{formlagr}
\mathcal{L}_A = -\frac{1}{48}F^2 + V(A^2),
\end{equation}
where squaring denotes contraction of all the indices, $A^2\equiv A^{\alpha\beta\gamma}A_{\alpha\beta\gamma}$, $V\equiv V(A^2)$ is the potential function, and the 4-form $\bf F=dA$ is the strength tensor \cite{Turok:1998he}, whose components can be written as,
\begin{equation}
\label{maxwell}
 F_{\alpha\beta\gamma\delta} = \nabla_{\alpha}A_{\beta\gamma\delta}-\nabla_{\delta}A_{\alpha\beta\gamma}+\nabla_{\gamma}A_{\delta\alpha\beta}-\nabla_{\beta}A_{\gamma\delta\alpha}.
\end{equation}
Computing the Euler-Lagrange equations we find that the equations of motion for our form field reads
\begin{equation}
\label{motion1}
\nabla_{\mu}F^{\mu\alpha\beta\gamma}=12\frac{\partial V}{\partial A^2}A^{\alpha\beta\gamma}.
\end{equation}

Theories employing three-form fields in cosmology hitherto are mostly applied to FLRW universes, e.g., in models of dark energy \cite{Morais:2016bev,Koivisto:2012xm,Koivisto:2009fb} and of inflation \cite{Koivisto:2009sd,Kumar:2016tdn,DeFelice:2012jt,DeFelice:2012wy,Barros:2015evi,Kumar:2014oka}, where they appear as a function of cosmic time $t$ (at the background level). However, since we are dealing with a static and spherically symmetric model, we are solely interested on a spatial dependence. One way of achieving this is by associating its components with a function, which we denote $\zeta(r)$, which is solely dependent on the radial coordinate. Due to the antisymmetric nature of the three-form, once $\zeta(r)$ is known, all of the other components are automatically determined.

It is common to write the 1-form (vector) $B^{\delta}$ \cite{Koivisto:2009sd}, dual to the three-form, via the Hodge star operator, $\star:\Omega^p(X)\rightarrow\Omega^{n-p}(X)$ (where $\Omega^p(X)$ is the vector space of $p$-forms on an $n$-dimensional smooth manifold $X$), which maps $p$-forms into $(n-p)$-forms, through the following relation:

\begin{equation}
\label{hodge1}
\left( \star A \right)^{\alpha_1 ... \alpha_{n-p}}= \frac{1}{p!}\frac{1}{\sqrt{-g}}\,\epsilon^{\alpha_1 ... \alpha_{n-p}\beta_1...\beta_{p}}A_{\beta_1...\beta_p},
\end{equation}
where $A$ is a $p-$form and $\epsilon$ is the $n-$dimensional Levi-Civita symbol. For our present study, where $n=4$ and $p=3$, Eq.~\eqref{hodge1} produces a vector,
\begin{equation}
\label{hodge}
B^{\delta} \equiv (\star A)^{\delta} = \frac{1}{3!}\frac{1}{\sqrt{-g}}\,\epsilon^{\delta\alpha\beta\gamma}A_{\alpha\beta\gamma}.
\end{equation}

We can now invert Eq.~\eqref{hodge} and express the components of the three-form in terms of its dual
\begin{equation}
\label{dual}
A_{\alpha\beta\gamma} = \sqrt{-g}\,\epsilon_{\alpha\beta\gamma\delta}B^{\delta}.
\end{equation}

We construct the components of the three-form by expressing the dual vector as a function of $\zeta (r)$, as
\begin{equation}
\label{duall}
B^{\delta} = \left( 0,\sqrt{1-\frac{b(r)}{r}}\,\zeta (r),0,0 \right)^{\rm T}.
\end{equation}
Note that $\zeta (r)$ is a convenient parametrization of the 3-form field in the geometry  given  by the line element (\ref{metric}).
From Eqs.~\eqref{dual} and \eqref{duall}, the non-zero components of the three-form read
\begin{eqnarray}
\label{components}
A_{t\theta\phi}&=&A_{\theta\phi t}=A_{\phi t\theta}=-A_{t\phi\theta}=-A_{\theta t\phi}=-A_{\phi\theta t} \nonumber\\
&=&\sqrt{g\left( \frac{b(r)}{r}-1 \right)}\,\zeta(r),
\end{eqnarray}
and $A^2$ is given by
\begin{equation}
A^2\equiv A^{\alpha\beta\gamma}A_{\alpha\beta\gamma}
= - 6\, \zeta(r)^2.
\end{equation}

Now, taking into account Eqs.~\eqref{maxwell} and \eqref{dual}, we can write the kinetic term in the Lagrangian Eq.~\eqref{formlagr} in a vector structure \cite{DeFelice:2012jt}
\begin{equation}
\label{ah}
-\frac{1}{48}F^2=-\frac{1}{2} F_{0123}F^{0123} = \frac{1}{2}(\nabla_{\mu}B^{\mu})^2,
\end{equation} 
and more specifically as 
\begin{equation}
F^2 = F^{\alpha\beta\gamma\delta}F_{\alpha\beta\gamma\delta}= - 6 \Upsilon
\end{equation}
where $\Upsilon$ is defined, for notational simplicity, by
\begin{equation}
\label{upsilon}
\Upsilon=4 \left( 1-\frac{b}{r} \right) \left[ \zeta\left( \Phi' + \frac{2}{r} \right) + \zeta' \right]^2.
\end{equation}
$\Upsilon$ can be tentatively interpreted as the kinetic energy of the three-form, which vanishes at the throat, $\Upsilon|_{r_0}=0$.

These relations will play an important role in solving the gravitational field equations, which is the aim of Section \ref{solutions}.

\subsection{Gravitational field equations}

Varying the action Eq.~\eqref{action} with respect to the metric $g_{\mu\nu}$, yields the gravitational field equations
\begin{equation}
\label{einsteineq}
G_{\mu\nu}=\kappa^2 \,T^{({\rm eff})}_{\mu\nu},
\end{equation}
where $G_{\mu\nu}$ is the Einstein tensor and we have defined the effective energy-momentum tensor of the sources by
\begin{equation}
T^{({\rm eff})}_{\mu\nu} = T^{({\rm A})}_{\mu\nu} + T^{(m)}_{\mu\nu}, 
\end{equation}
where an $({\rm A})$ superscript refers to the three-form $A_{\alpha\beta\gamma}$ and an $(m)$ to other matter sources. Note that the energy-momentum tensor associated with the $i$-species is defined as
\begin{equation}
\label{emgeral}
T^{\rm (i)}_{\mu\nu} = -2 \frac{\delta \mathcal{L}_{\rm (i)}}{\delta g^{\mu\nu}}+g_{\mu\nu}\mathcal{L}_{\rm (i)}.
\end{equation}

Taking into account Eqs.~\eqref{formlagr} and \eqref{emgeral}, we find that the energy-momentum tensor of the three-form reads
\begin{equation}
T^{({\rm A})}_{\mu\nu} = \frac{1}{6}F_{\mu\alpha\beta\gamma}F_{\nu}\,^{\alpha\beta\gamma} +6\frac{\partial V}{\partial A^2}A_{\mu\alpha\beta}A_{\nu}\,^{\alpha\beta}+ \mathcal{L}_A\,g_{\mu\nu},
\label{EMT3form}
\end{equation}
in which case, taking into account the wormhole metric \eqref{metric}, its components are given by
\begin{eqnarray}
T^{{\rm (A)}}\,^t_t &=& - \rho_{\rm A}= -\frac{1}{8}\Upsilon - V + \zeta V_{,\zeta}, 
\\
T^{\rm (A)}\,^r_r &=& - \tau_{\rm A}= -\frac{1}{8}\Upsilon - V, 
\\
T^{\rm (A)}\,^{\theta}_{\theta} &=& p_{\rm A}= T^{\rm (A)}\,^{\phi}_{\phi} = -\frac{1}{8}\Upsilon - V + \zeta V_{,\zeta}.
\end{eqnarray} 

Note that in a cosmological context, it is already known that \cite{Koivisto:2012xm,Koivisto:2009fb}, in the absence of a potential in the Lagrangian Eq.~\eqref{formlagr}, the three-form source mimics a cosmological constant with an equation of state $w=p_{\rm A}/\rho_{\rm A}=-1$. This fact was used by Hawking  \cite{Turok:1998he} to tackle the cosmological constant problem.

Now, consider that the matter energy-momentum tensor consists of an anisotropic fluid, given by $T^{\mu}{}_{\nu}={\rm diag}(-\rho_m,-\tau_m,p_m,p_m)$, where $\rho_m$ is the matter energy density, $\tau_m$ is the radial tension, and $p_m$ is the tangential pressure of matter.

With these definitions, the gravitational field equations \eqref{einsteineq} yield
\begin{eqnarray}
\rho_{{\rm eff}} &=& \rho_{m}  + \rho_{{\rm A}} =
 \frac{b'}{r^2}, \label{ee1}
 	\\
\tau_{{\rm eff}} &=& \tau_{m}  + \tau_{{\rm A}}=
 \frac{b}{r^3} - 2\left( 1-\frac{b}{r} \right)\frac{\Phi'}{r}, \label{ee2} \\
p_{{\rm eff}} &=& p_{m} + p_{{\rm A}} = \left( 1-\frac{b}{r}\right)\left[ \Phi''+\Phi'^2 - \frac{b'r-b}{2r(r-b)}\Phi' \right. 
	\nonumber \\
& &\left. - \frac{b'r-b}{2r^2(r-b)}+\frac{\Phi'}{r} \right],\label{ee3}
\end{eqnarray}
where for simplicity, we have set $\kappa^2 = 1$, and omitted the parameter dependencies, e.g., $\Phi\equiv\Phi(r)$; we follow this notation throughout this work. We can interpret the components of the effective energy-momentum tensor as $\rho_{{\rm eff}}(r)$ being the total energy density, $\tau_{{\rm eff}}(r)$ the radial tension (negative of the radial pressure) and $p_{{\rm eff}}(r)$ the pressure on the tangential directions, orthogonal to the radial direction.

The Bianchi identities yield the conservation of the effective energy-momentum tensor, i.e., $\nabla_{\mu}T^{({\rm eff})}\,^{\mu}_{\nu} = 0$, in which the radial component imposes the following continuity equation:
\begin{equation}
\tau_{{\rm eff}}'+\frac{2}{r}\left( \tau_{{\rm eff}} + p_{{\rm eff}} \right) + \Phi' \left( \tau_{{\rm eff}} - \rho_{{\rm eff}} \right)=0.
\end{equation}

The equations of motion, Eq.~\eqref{motion1}, can now be written in terms of the scalar function $\zeta$ as
\begin{eqnarray}
&\zeta \left[ r\,\Phi' \left( b' - \frac{b}{r} \right) +4 + 2\,b' + 2\,r^2\, \Phi'' \left( \frac{b}{r}-1 \right) -6\,\frac{b}{r} \right]  \nonumber \\
&+ 2r^2 V_{,\zeta}+ \zeta'r\left[ 3\,\frac{b}{r} - 4 + b' + 2\Phi'\, r \left( \frac{b}{r}-1 \right) \right] \nonumber \\
&+2\,\zeta''r^2 \left( \frac{b}{r}-1 \right) = 0 \,. \label{eqmotzeta}
\end{eqnarray}
This relation places a further constraint on the unknown functions, and will play a major role in finding explicit wormhole solutions.

\section{Energy conditions}\label{ec}

Now, as mentioned above, the fundamental ingredient in wormhole physics is the flaring-out condition, which through the Einstein field equations, in standard GR, entails the violation of the NEC. The latter states that, for every future oriented null vector $k^{\mu}$, the relation $T_{\mu\nu}k^{\mu}k^{\nu}\geq 0$ holds. However, relative to this issue, we need to point out a subtlety. The energy conditions arise when one refers back to the Raychaudhuri equation, where a geometric term $R_{\mu\nu}k^\mu k^\nu$ appears. Note that one imposes the positivity of this term, i.e., $R_{\mu\nu}k^\mu k^\nu \geq 0$, to ensure the focussing of geodesic congruences, within a finite value of the parameter labelling the points on the geodesics. However, through the Einstein field equations, the positivity condition entails the condition $T_{\mu\nu}k^\mu k^\nu \geq 0$ on the energy-momentum tensor, which is simply translated as the NEC.

In the present theory, with three-forms, things are not so straightforward, where the effective Einstein field equations, given by Eq.~\eqref{einsteineq}, may now be rewritten as $R_{\mu\nu}=\kappa^2 \left(T^{(\rm eff)}_{\mu\nu} - \frac{1}{2}T^{(\rm eff)}g_{\mu\nu}\right)$. Here, one may replace the Ricci tensor $R_{\mu\nu}$ by the corresponding field equation with the matters sources. This argument shows that now it is the effective energy-momentum tensor that violates the NEC, i.e., $T^{(\rm eff)}_{\mu\nu} k^\mu k^\nu < 0$. Thus, in principle, one may consider that the matter energy-momentum tensor $T^{(m)}_{\mu\nu}$ satisfies the energy conditions and the respective violations arise from the three-form curvature term $T^{(\rm A)}_{\mu\nu}$. In fact, it is useful to impose the condition $T^{(m)}_{\mu\nu} k^\mu k^\nu\geq 0$ at face value, as taking into account local Lorentz transformations one may show that this condition implies the positivity of the energy density in all local frames of reference. Indeed, the flaring-out condition, which entails $R_{\mu\nu}\,k^\mu k^\nu < 0$ in the vicinity of the throat does not imply the focussing of geodesics, which  translates a repulsive character of gravity in this region.

Therefore, the strategy in this paper will be the following. We impose that the matter energy-momentum tensor does indeed satisfy the condition $T^{(m)}_{\mu\nu} k^\mu k^\nu\geq 0$, and it is the three-form curvature term that entails the violation of the \cite{Capozziello:2018wul,Capozziello:2013vna,Capozziello:2014bqa} ``generalized'' NEC \citep{Harko:2013yb}, i.e., $T^{(\rm eff)}_{\mu\nu} k^\mu k^\nu < 0$. This implies the following constraint:
\begin{equation}
0 \leq T^{(m)}_{\mu\nu}k^{\mu}k^{\nu} < - T^{({\rm A})}_{\mu\nu}k^{\mu}k^{\nu},
\end{equation}
in the vicinity of the throat, where taking into account the energy-momentum of the three-form, i.e., Eq.~\eqref{EMT3form}, one deduces the fundamental inequality in this work, given by
\begin{eqnarray}
0 \leq T^{(m)}_{\mu\nu}k^{\mu}k^{\nu} &<& - \frac{1}{6}F_{\mu\alpha\beta\gamma}F_{\nu}\,^{\alpha\beta\gamma}\, k^{\mu}k^{\nu} 
	\nonumber  \\
&& -6\frac{\partial V}{\partial A^2}A_{\mu\alpha\beta}A_{\nu}\,^{\alpha\beta}\, k^{\mu}k^{\nu}  \,.
\end{eqnarray}

Now, considering that the world line of an observer (or family of observers) can be expressed through a timelike vector $V^{\mu}$, the weak energy condition (WEC) imposes that the energy density measured by this observer is always non-negative, i.e., $T_{\mu\nu}V^{\mu}V^{\nu} \geq 0$. In the present case, we impose that the energy density of the matter threading the wormhole satisfies the WEC, so that 
\begin{eqnarray}
T^{(m)}_{\mu\nu}\,V^{\mu}V^{\nu} =  \left(G_{\mu\nu}/\kappa^2  -  T^{({\rm A})}_{\mu\nu} \right) \,V^{\mu}V^{\nu}   \geq 0 \,,
\end{eqnarray}
and using Eq.~\eqref{EMT3form}, takes the following form
\begin{eqnarray}
T^{(m)}_{\mu\nu}\,V^{\mu}V^{\nu} &=& \frac{1}{\kappa^2} G_{\mu\nu} \, V^{\mu}V^{\nu}  - \frac{1}{6}F_{\mu\alpha\beta\gamma}F_{\nu}\,^{\alpha\beta\gamma}\, V^{\mu}V^{\nu} 
	\nonumber  \\
&-&6\frac{\partial V}{\partial A^2}A_{\mu\alpha\beta}A_{\nu}\,^{\alpha\beta}\, V^{\mu}V^{\nu}  +  \mathcal{L}_A \geq 0  .
\end{eqnarray}

In summary, for the matter energy-momentum tensor with components $T^{\mu}{}_{\nu}={\rm diag}(-\rho_m,-\tau_m,p_m,p_m)$ (where $\tau_m=-p^r_m$, i.e., the matter radial tension equals the negative matter radial pressure), the flaring-out condition, i.e., $b'(r) < b(r)/r$, entails the following inequality on the effective energy-momentum tensor components, $\rho_{\rm eff}-\tau_{\rm eff}  < 0$. However, in this work, we are interested in imposing that the matter energy-momentum tensor $T^{(m)}_{\mu\nu}$ threading the wormhole satisfies the WEC throughout the entire spacetime, which imposes the following inequalities 
\begin{equation}
\rho_{m} \geq 0 \qquad {\rm and} \qquad \rho_{m}-\tau_{m} \geq 0 \,.
\end{equation}
Thus, it is the energy-momentum curvature term associated to the three-form that is responsible for sustaining these wormhole geometries.

\begin{figure*}[htp]
	\begin{center}
		\includegraphics[width=.45\textwidth]{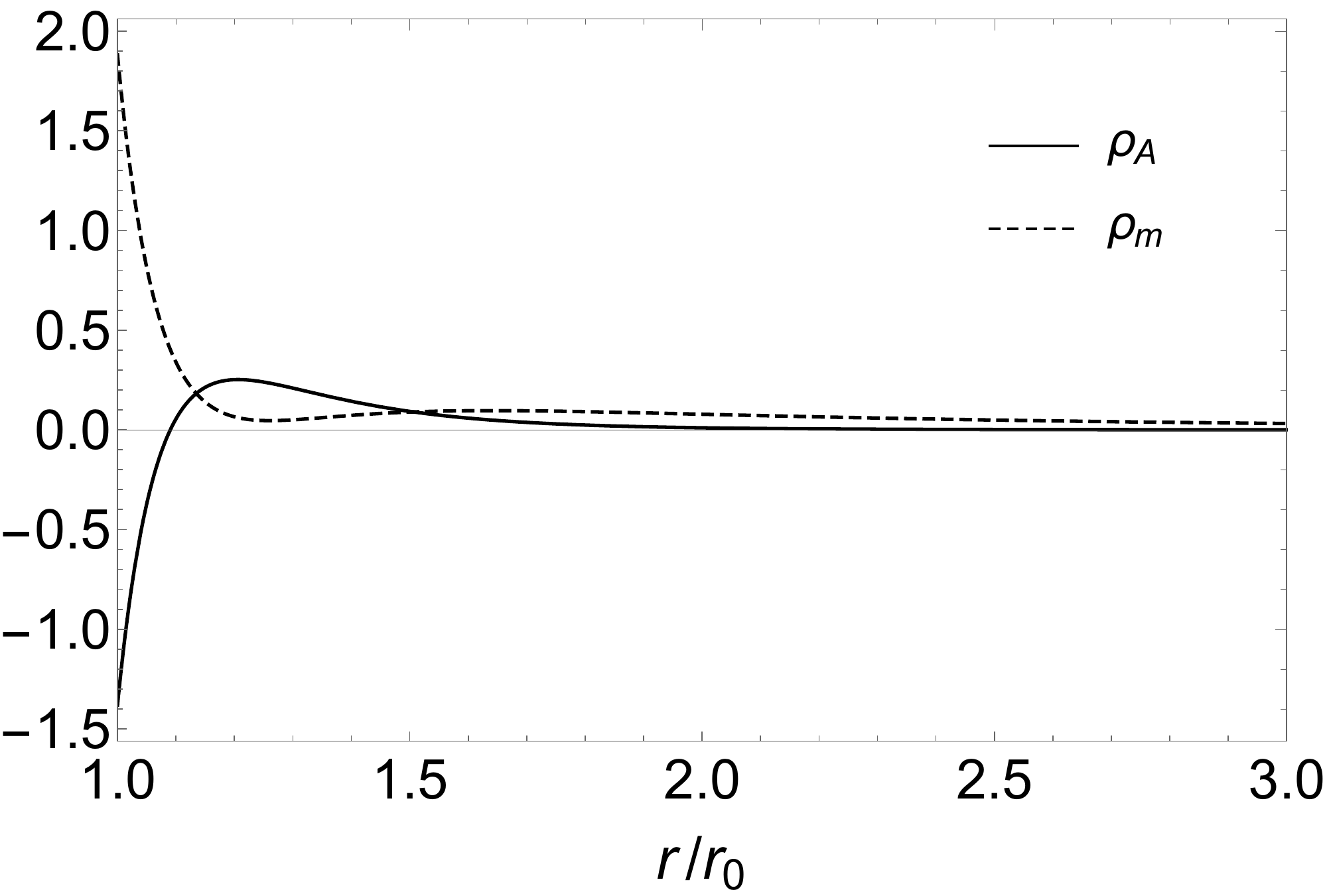}
		\hfill
		\includegraphics[width=.45\textwidth]{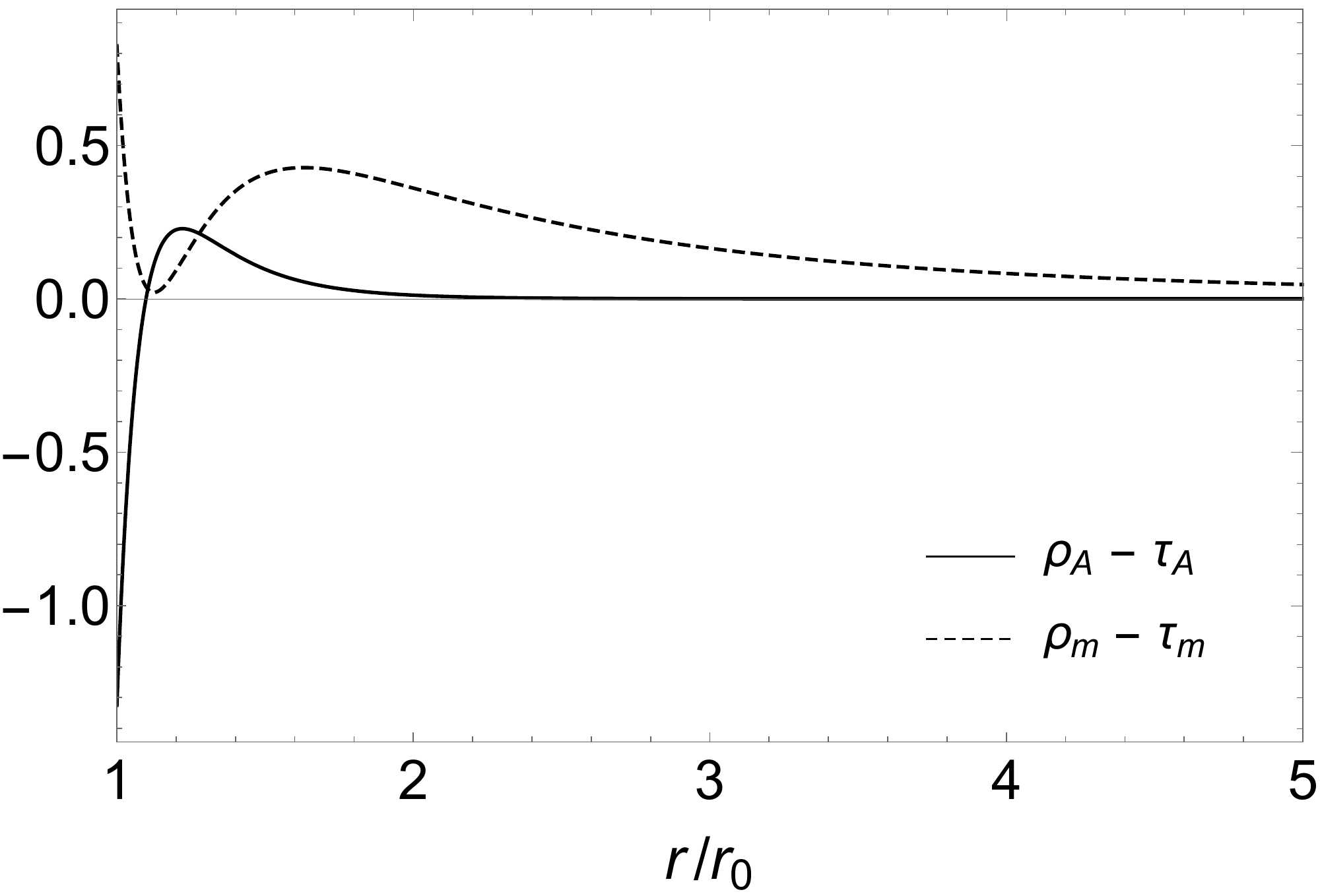}
	\end{center}
	\caption{\label{fig1}Energy densities (left panel) and NEC profile (right panel) for the form field (solid) and for the matter sources (dashed), regarding the specific choice given by Eqs.~\eqref{sol1}-\eqref{sol1b} with $\beta=-1/2$, $\Phi_0 = -6.3$, $\alpha = 1$, $\zeta_0 = 1$, $\gamma = 3$ and $C=0$. We refer the reader to the text for more details.}
\end{figure*}

\section{Specific solutions}\label{solutions}

In order to find wormhole solutions, we need to solve the four independent equations, which consist of three gravitational field equations Eqs.~\eqref{ee1}--\eqref{ee3} and the equation of motion for $\zeta$, i.e., Eq.~\eqref{eqmotzeta}. The system has seven unknown functions, namely, $\Phi$, $b$, $\rho_m$, $\tau_m$, $p_m$, $\zeta$ and $V$, so we have up to three assumptions to make. As has been done in previous works \cite{Lobo:2008zu,Lobo:2005us,Capozziello:2012hr}, we will specify the redshift and shape functions, and assume further a form for $\zeta$ (which we call type I, see subsection \ref{typeI}) or for the potential (type II, see subsection \ref{typeII}) to solve the system. For the cases where the potential is specified and non constant, we are unable to find analytical solutions, so we consider a numerical analysis to integrate the equations by specifying the initial conditions far from the throat, and integrate to the throat.

\subsection{Type I solutions}\label{typeI}

Following the notation of \cite{Capozziello:2012hr}, we consider the following choices for the metric functions
\begin{equation}
\label{sol1}
b(r)=r_0\left( \frac{r_0}{r} \right)^{\beta}, \qquad  \Phi(r)=\Phi_0\left( \frac{r_0}{r} \right)^{\alpha},
\end{equation}
and for the $\zeta$ function
\begin{equation}
\label{sol1b}
\zeta(r)=\zeta_0\left( \frac{r_0}{r} \right)^{\gamma},
\end{equation}
where $\beta>-1$, $\alpha>0$ and $\gamma>0$. Note that Eq.~\eqref{sol1b} takes the value $\zeta=\zeta_0$ at the throat and tends to zero at spatial infinity. 

Substituting the specific choices \eqref{sol1} and \eqref{sol1b} into Eq.~\eqref{eqmotzeta}, the latter becomes a first order differential equation for $V(r)$ where we find the following analytical solution,
\begin{eqnarray}
V &=& \frac{\zeta_0^2 \gamma}{2r^2} \left\{ \left[ 1-\left( \frac{r_0}{r} \right)^{\beta+1}\right] \left( \gamma -2 \right) \right. \nonumber \\
&& + \Phi_0 \,\alpha \left( \frac{r_0}{r} \right)^{\alpha}\left[ 1 + \frac{\alpha}{\alpha + 2\left( 1+\gamma \right)} \right. \nonumber \\
&& -\left.\left. \left( \frac{r_0}{r} \right)^{\beta+1} \frac{3 + \beta + 2 \left( \alpha+\gamma \right)}{3+\beta+\alpha+2\gamma}\right]\right\}+C,
	\label{solV1}
\end{eqnarray}
in which $C$ is a constant.

In Fig.~\ref{fig1} we show the energy densities (left panel) and the NEC profile (right panel) of a particular solution where the matter component does not violate the NEC nor the WEC. This means that the three-form field is responsible for sustaining the wormhole, and all the exoticity of the object is confined to the field itself and the matter sources thread the wormhole without violating the NEC and WEC. This is the main virtue of these models.

\begin{figure*}[htp]
	\begin{center}
		\includegraphics[width=0.45\textwidth]{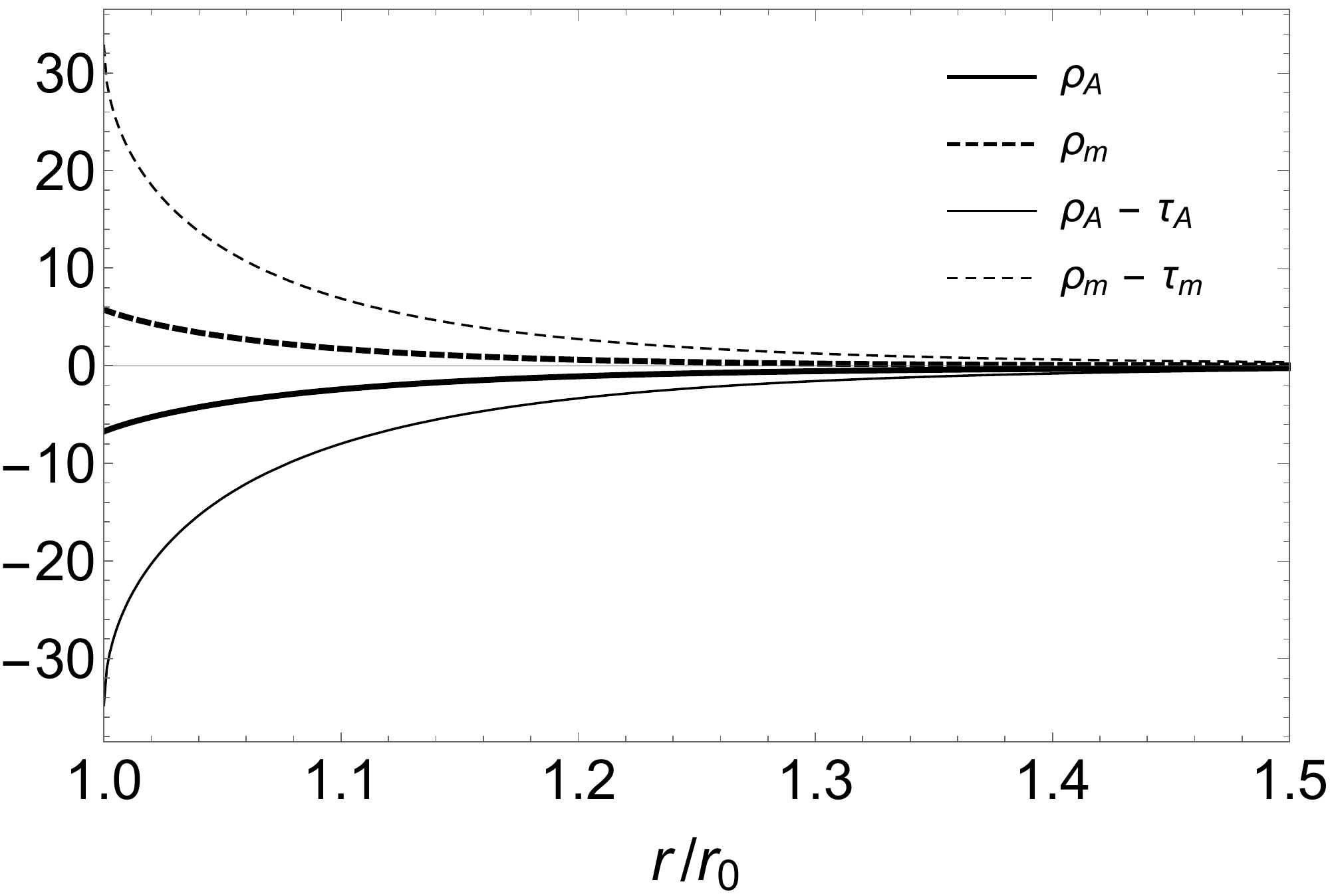}
		\hfill
		\includegraphics[width=0.45\textwidth]{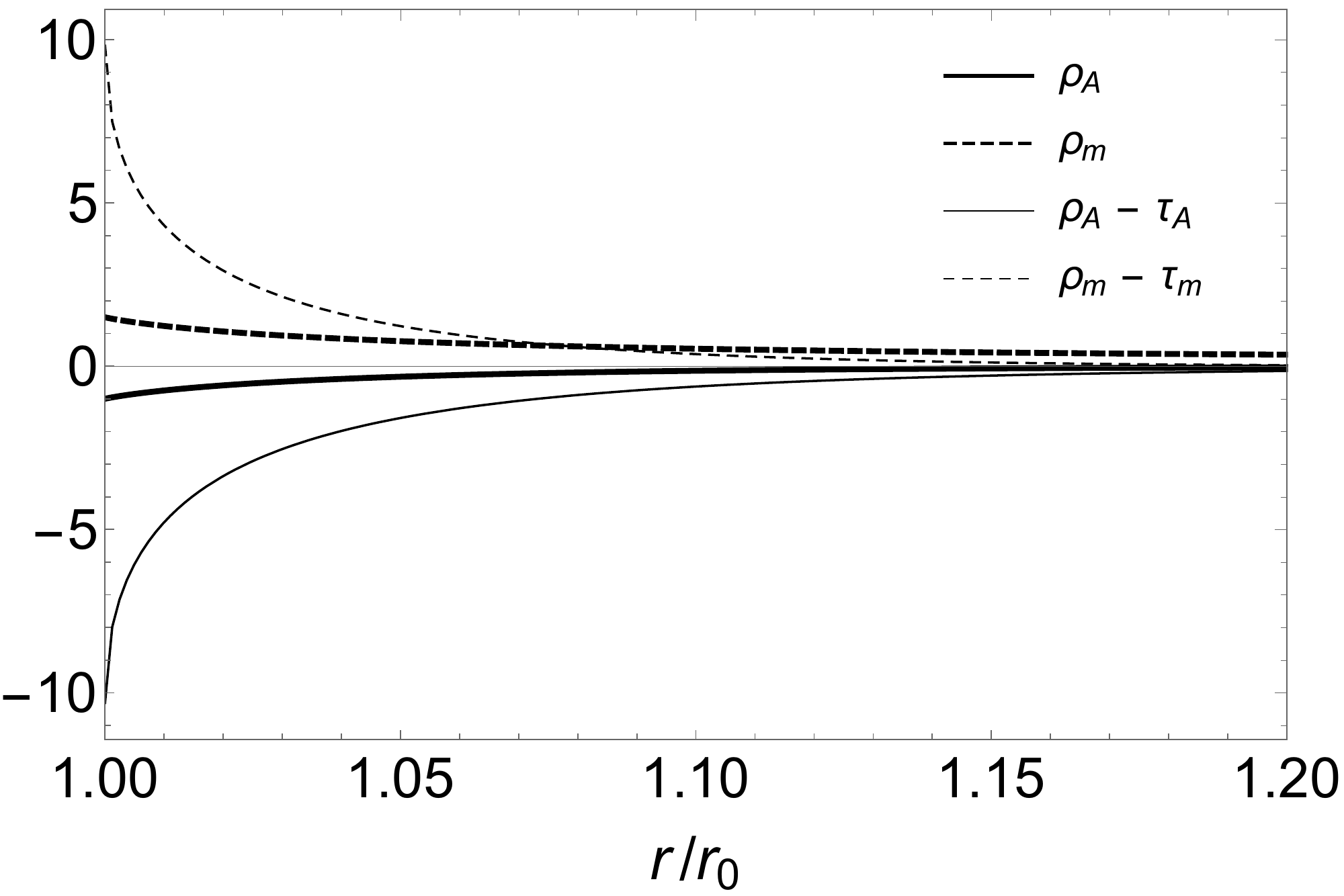}
	\end{center}
	\caption{\label{fig2}Energy densities (thick) and NEC profiles (thin) for the form field (solid) and the matter sources (dashed) for the choices in Eqs.~\eqref{sol1}-\eqref{sol1b} and Eq.~\eqref{sol2} and . Left panel: $\beta = 1$, $\Phi_0 = -1$, $\alpha=1$ and $C=-0.1$. Right panel: $\beta = -1/2$, $\Phi_0 = -2$, $\alpha=1$ and $C=0$. See the text for more details.}
\end{figure*}

An interesting case is obtained by considering a zero redshift function $\Phi_0=0$ and $\gamma = 2$, so that the potential becomes a constant $V=C$, which is readily found from Eq.~\eqref{solV1}. In this particular case, although $\zeta$ is dependent on the radial coordinate, we note that its kinetic term in Eq.~\eqref{upsilon} vanishes, the energy density of the form field is constant, and is given by $\rho_{\rm A}=V=C$. This means that the field mimics a cosmological constant \cite{Lemos:2003jb}. This feature does not happen in general with classical scalar fields when they are not constant. It is due to the fact that the kinetic term of the energy density of the three-form, Eq.~\eqref{upsilon}, depends on $\zeta'$ and $\zeta$ itself (in contrast with classical canonical scalar fields where the kinetic term depends solely on the field derivative and not explicitly on the field itself). Thus, the $\zeta'$ and $\zeta$ terms can mutually cancel. If $C\geqslant 0$ the field does not violate the NEC and WEC, however the matter fields in this case are exotic, which is not our main interest in this study. This feature tells us that it is also possible for a three-form to exist within a wormhole without violating the energy conditions while mimicking a cosmological constant.

\subsection{Type II solutions}\label{typeII}

We now consider the metric functions given by \eqref{sol1} and specify a potential $V$ with a quadratic form,
\begin{equation}
\label{sol2} V(\zeta)=\zeta^2 + C.
\end{equation}
%
The equation of motion \eqref{eqmotzeta} now becomes a second order differential equation for $\zeta$ for which we are not able to find analytical solutions, thus we resort to a numerical analysis. The results are reported in  Fig. \ref{fig2}, for $b = r_0^2/r$ (left panel) and $b=\sqrt{r_0 r}$ (right panel), for different choices for the constants (see caption). 
We show that it is possible to recreate a similar behavior, where the matter fields do not violate the NEC nor the WEC and all the exoticity is contained in the field itself. 

\begin{figure}[htp]
	\begin{center}
		\includegraphics[width=0.45\textwidth]{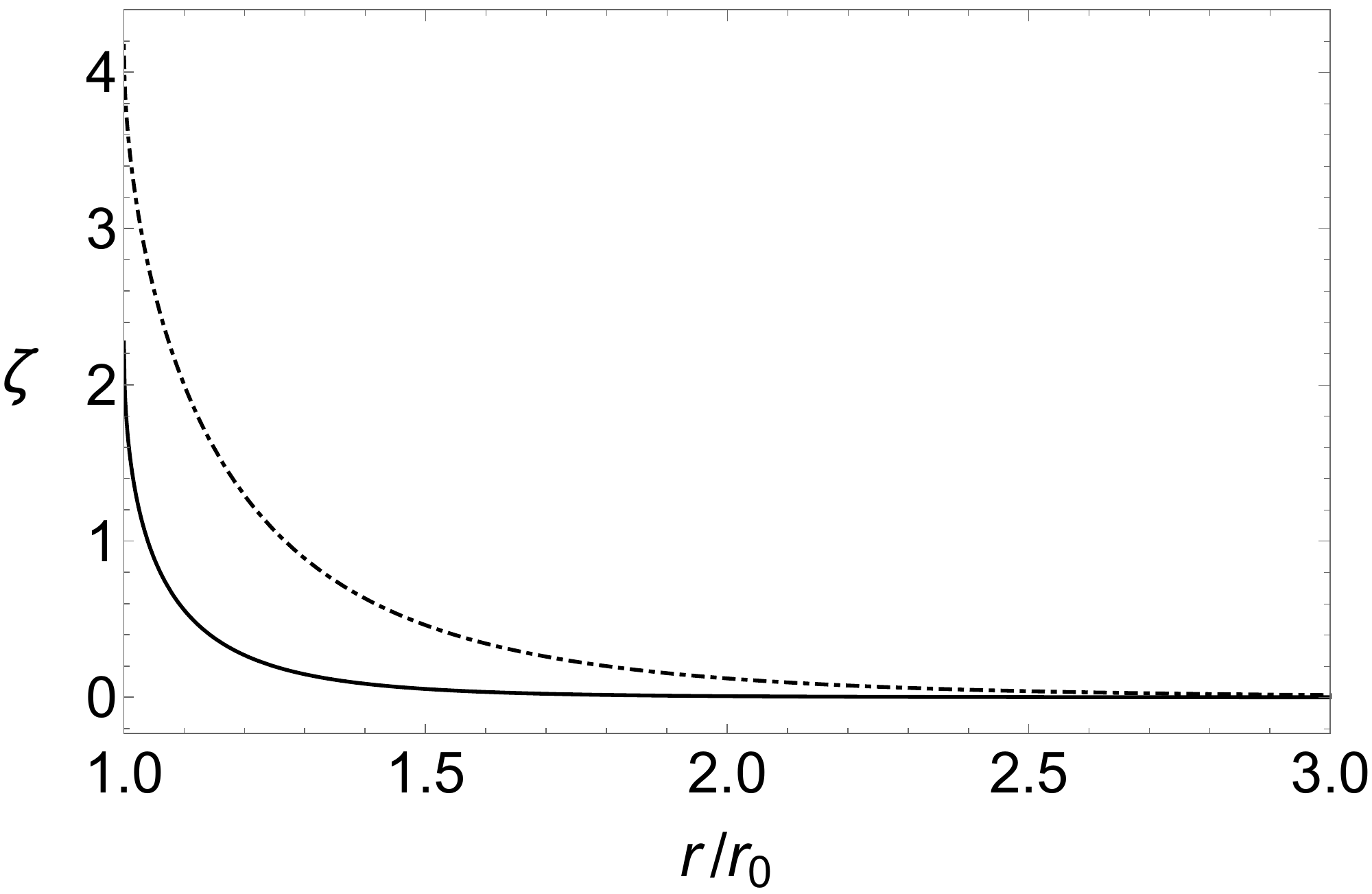}
	\end{center}
	\caption{\label{zeta}Solutions for $\zeta (r)$, regarding the two solutions of Eqs.~\eqref{sol1}-\eqref{sol1b} and Eq.~\eqref{sol2} for: $\beta = -1/2$, $\Phi_0 = -2$, $\alpha=1$ and $C=0$ (solid) and $\beta = 1$, $\Phi_0 = -1$, $\alpha=1$ and $C=-0.1$ (dot dashed). See the text for details.}
\end{figure}

The solutions for $\zeta$ are displayed in Fig. \ref{zeta}. We observe that near the throat the function takes non-zero values and smoothly decays as the wormhole flares, vanishing as $r\rightarrow \infty$. We also note that, for the case where $b=\sqrt{r_0 r}$, $\zeta$ decays faster at the vicinity of $r_0$, therefore, in this setting, the wormhole will cluster the energy densities closer to its throat.

It is interesting to consider the particular case of the zero tidal force, i.e., $\Phi(r)=0$, taking into account that the shape function follows the form $b(r) = r_0^2/r$ and a constant potential $V=V_0$, Eq.~\eqref{eqmotzeta} finally becomes
\begin{gather}
\zeta'' r^2 \left[ \left(\frac{r}{r_0}\right)^2-1\right] + \zeta' r \left[ 2 \left( \frac{r}{r_0}\right)^2 -1 \right] \nonumber \\
- 2\zeta \left[ \left( \frac{r}{r_0}\right)^2 -2\right] =0,
\end{gather} 
which yields the following analytical solution
\begin{equation}
\zeta (r)= \frac{C_1}{r^2} + C_2 \sqrt{\left( \frac{r}{r_0}\right)^2 -1} \left[1+2\left( \frac{r_0}{r}\right)^2\right], 
\label{zetasolution}
\end{equation}
where $C_1$ and $C_2$ are constants. In this framework, the energy density of the form field is constant and reads
\begin{equation}
\label{an}
\rho_{\textup{A}}=\frac{9}{2}\left(\frac{C_2}{r_0}\right)^2 + V_0.
\end{equation}

Note that Eq.~\eqref{an} is the same of the specific one found in \ref{typeI} when $C_2=0$. However, in the previous setting where the shape of $\zeta$ was assumed, the energy density depended only on $V_0$. In this case, we find a generalized form for $\zeta$ which can also mimic a cosmological constant in the absence of a potential, $V_0=0$, when $C_2\neq 0$.
However, in order to avoid divergences, we impose that $C_2=0$, so that $\zeta$ in Eq. (\ref{zetasolution}) tends to zero and spatial infinity.

\section{Conclusions}\label{conclusions}

In this work, it was shown that it is possible to sustain static and spherically symmetric wormhole geometries with a three-form field. More specifically, it was shown that it is possible to find analytical and numerical solutions to the gravitational field equations, where the matter fields threading the wormhole do not violate the null nor the weak energy conditions and all the exoticity is contained in the three-form field. An interesting case was also presented where the three-form field mimics a cosmological constant, with a constant energy density even though the field varies with the radial coordinate, which is in contrast with classical scalar fields. It was also shown, although not as appealing, that it is possible to find solutions where the form field does not violate the NEC and WEC, and it is the matter fields that sustain the wormhole geometry, and consequently violates the WEC. Thus, it was shown that three-form curvature terms, which may be interpreted as a gravitational fluid, sustain these wormhole geometries \cite{Harko:2013yb}.

We emphasize that the three-form application to these compact objects is a novel approach to wormhole physics, which opens new avenues of research. As mentioned in the Introduction, the stability of these wormhole geometries is of great importance, but lies outside the scope of this work. However, a future line of research lies in applying a similar analysis as considered in \cite{Kleihaus:2017kai,Kleihaus:2014dla} to these solutions using three-forms. Another application is to construct evolving wormhole geometries, such as finding solutions in a cosmological background, which are conformally related to the respective static geometries,
as outlined in \cite{Kar:1995ss,Arellano:2006ex,Lobo:2007zb,Bronnikov:2017sgg}. It was shown that these dynamical geometries exhibit flashes of time satisfying the WEC. Thus, we also propose to analyze evolving wormhole geometries, constructed by dynamical three-forms and investigate the possibility that these do in fact satisfy the WEC in specific regions of time. Work along these lines is presently underway.

\section{Acknowledgements}
BJB is supported by Funda\c{c}\~ao para a Ci\^encia e a Tecnologia (FCT, Portugal) through the grant PD/BD/128018/2016.
FSNL is funded through an Investigador FCT Research contract No.~IF/00859/2012. BJB and FSNL also acknowledge funding from the research grants UID/FIS/04434/2013 and No.~PEst-OE/FIS/UI2751/2014.

\bibliography{bib1}

\begin{thebibliography}{53}%
\makeatletter
\providecommand \@ifxundefined [1]{%
 \@ifx{#1\undefined}
}%
\providecommand \@ifnum [1]{%
 \ifnum #1\expandafter \@firstoftwo
 \else \expandafter \@secondoftwo
 \fi
}%
\providecommand \@ifx [1]{%
 \ifx #1\expandafter \@firstoftwo
 \else \expandafter \@secondoftwo
 \fi
}%
\providecommand \natexlab [1]{#1}%
\providecommand \enquote  [1]{``#1''}%
\providecommand \bibnamefont  [1]{#1}%
\providecommand \bibfnamefont [1]{#1}%
\providecommand \citenamefont [1]{#1}%
\providecommand \href@noop [0]{\@secondoftwo}%
\providecommand \href [0]{\begingroup \@sanitize@url \@href}%
\providecommand \@href[1]{\@@startlink{#1}\@@href}%
\providecommand \@@href[1]{\endgroup#1\@@endlink}%
\providecommand \@sanitize@url [0]{\catcode `\\12\catcode `\$12\catcode
  `\&12\catcode `\#12\catcode `\^12\catcode `\_12\catcode `\%12\relax}%
\providecommand \@@startlink[1]{}%
\providecommand \@@endlink[0]{}%
\providecommand \url  [0]{\begingroup\@sanitize@url \@url }%
\providecommand \@url [1]{\endgroup\@href {#1}{\urlprefix }}%
\providecommand \urlprefix  [0]{URL }%
\providecommand \Eprint [0]{\href }%
\providecommand \doibase [0]{http://dx.doi.org/}%
\providecommand \selectlanguage [0]{\@gobble}%
\providecommand \bibinfo  [0]{\@secondoftwo}%
\providecommand \bibfield  [0]{\@secondoftwo}%
\providecommand \translation [1]{[#1]}%
\providecommand \BibitemOpen [0]{}%
\providecommand \bibitemStop [0]{}%
\providecommand \bibitemNoStop [0]{.\EOS\space}%
\providecommand \EOS [0]{\spacefactor3000\relax}%
\providecommand \BibitemShut  [1]{\csname bibitem#1\endcsname}%
\let\auto@bib@innerbib\@empty
\bibitem [{\citenamefont {Morris}\ and\ \citenamefont
  {Thorne}(1988)}]{Morris:1988cz}%
  \BibitemOpen
  \bibfield  {author} {\bibinfo {author} {\bibfnamefont {M.~S.}\ \bibnamefont
  {Morris}}\ and\ \bibinfo {author} {\bibfnamefont {K.~S.}\ \bibnamefont
  {Thorne}},\ }\href {\doibase 10.1119/1.15620} {\bibfield  {journal} {\bibinfo
   {journal} {Am. J. Phys.}\ }\textbf {\bibinfo {volume} {56}},\ \bibinfo
  {pages} {395} (\bibinfo {year} {1988})}\BibitemShut {NoStop}%
\bibitem [{\citenamefont {Visser}(1995)}]{Visser:1995cc}%
  \BibitemOpen
  \bibfield  {author} {\bibinfo {author} {\bibfnamefont {M.}~\bibnamefont
  {Visser}},\ }\href@noop {} {\emph {\bibinfo {title} {{Lorentzian wormholes:
  From Einstein to Hawking}}}}\ (\bibinfo {year} {1995})\BibitemShut {NoStop}%
\bibitem [{\citenamefont {Flamm}(1916)}]{Flamm}%
  \BibitemOpen
  \bibfield  {author} {\bibinfo {author} {\bibfnamefont {L.}~\bibnamefont
  {Flamm}},\ }\href@noop {} {\bibfield  {journal} {\bibinfo  {journal} {Phys.
  Z.}\ }\textbf {\bibinfo {volume} {17}} (\bibinfo {year} {1916})}\BibitemShut
  {NoStop}%
\bibitem [{\citenamefont {Einstein}\ and\ \citenamefont
  {Rosen}(1935)}]{PhysRev.48.73}%
  \BibitemOpen
  \bibfield  {author} {\bibinfo {author} {\bibfnamefont {A.}~\bibnamefont
  {Einstein}}\ and\ \bibinfo {author} {\bibfnamefont {N.}~\bibnamefont
  {Rosen}},\ }\href {\doibase 10.1103/PhysRev.48.73} {\bibfield  {journal}
  {\bibinfo  {journal} {Phys. Rev.}\ }\textbf {\bibinfo {volume} {48}},\
  \bibinfo {pages} {73} (\bibinfo {year} {1935})}\BibitemShut {NoStop}%
\bibitem [{\citenamefont {Wheeler}(1955)}]{Wheeler:1955zz}%
  \BibitemOpen
  \bibfield  {author} {\bibinfo {author} {\bibfnamefont {J.~A.}\ \bibnamefont
  {Wheeler}},\ }\href {\doibase 10.1103/PhysRev.97.511} {\bibfield  {journal}
  {\bibinfo  {journal} {Phys. Rev.}\ }\textbf {\bibinfo {volume} {97}},\
  \bibinfo {pages} {511} (\bibinfo {year} {1955})}\BibitemShut {NoStop}%
\bibitem [{\citenamefont {Misner}\ and\ \citenamefont
  {Wheeler}(1957)}]{Misner:1957mt}%
  \BibitemOpen
  \bibfield  {author} {\bibinfo {author} {\bibfnamefont {C.~W.}\ \bibnamefont
  {Misner}}\ and\ \bibinfo {author} {\bibfnamefont {J.~A.}\ \bibnamefont
  {Wheeler}},\ }\href {\doibase 10.1016/0003-4916(57)90049-0} {\bibfield
  {journal} {\bibinfo  {journal} {Annals Phys.}\ }\textbf {\bibinfo {volume}
  {2}},\ \bibinfo {pages} {525} (\bibinfo {year} {1957})}\BibitemShut {NoStop}%
\bibitem [{\citenamefont {Ellis}(1973)}]{Ellis:1973yv}%
  \BibitemOpen
  \bibfield  {author} {\bibinfo {author} {\bibfnamefont {H.~G.}\ \bibnamefont
  {Ellis}},\ }\href {\doibase 10.1063/1.1666161} {\bibfield  {journal}
  {\bibinfo  {journal} {J. Math. Phys.}\ }\textbf {\bibinfo {volume} {14}},\
  \bibinfo {pages} {104} (\bibinfo {year} {1973})}\BibitemShut {NoStop}%
\bibitem [{\citenamefont {Bronnikov}(1973)}]{Bronnikov:1973fh}%
  \BibitemOpen
  \bibfield  {author} {\bibinfo {author} {\bibfnamefont {K.~A.}\ \bibnamefont
  {Bronnikov}},\ }\href@noop {} {\bibfield  {journal} {\bibinfo  {journal}
  {Acta Phys. Polon.}\ }\textbf {\bibinfo {volume} {B4}},\ \bibinfo {pages}
  {251} (\bibinfo {year} {1973})}\BibitemShut {NoStop}%
\bibitem [{\citenamefont {Lobo}(2005{\natexlab{a}})}]{Lobo:2004rp}%
  \BibitemOpen
  \bibfield  {author} {\bibinfo {author} {\bibfnamefont {F.~S.~N.}\
  \bibnamefont {Lobo}},\ }\href {\doibase 10.1007/s10714-005-0177-x} {\bibfield
   {journal} {\bibinfo  {journal} {Gen. Rel. Grav.}\ }\textbf {\bibinfo
  {volume} {37}},\ \bibinfo {pages} {2023} (\bibinfo {year}
  {2005}{\natexlab{a}})},\ \Eprint {http://arxiv.org/abs/gr-qc/0410087}
  {arXiv:gr-qc/0410087 [gr-qc]} \BibitemShut {NoStop}%
\bibitem [{\citenamefont {Curiel}(2017)}]{Curiel:2014zba}%
  \BibitemOpen
  \bibfield  {author} {\bibinfo {author} {\bibfnamefont {E.}~\bibnamefont
  {Curiel}},\ }\href {\doibase 10.1007/978-1-4939-3210-8_3} {\bibfield
  {journal} {\bibinfo  {journal} {Einstein Stud.}\ }\textbf {\bibinfo {volume}
  {13}},\ \bibinfo {pages} {43} (\bibinfo {year} {2017})},\ \Eprint
  {http://arxiv.org/abs/1405.0403} {arXiv:1405.0403 [physics.hist-ph]}
  \BibitemShut {NoStop}%
\bibitem [{\citenamefont {Lobo}(2017)}]{Lobo:2017oab}%
  \BibitemOpen
  \bibfield  {author} {\bibinfo {author} {\bibfnamefont {F.~S.~N.}\
  \bibnamefont {Lobo}},\ }\href {\doibase 10.1007/978-3-319-55182-1} {\bibfield
   {journal} {\bibinfo  {journal} {Fundam. Theor. Phys.}\ }\textbf {\bibinfo
  {volume} {189}},\ \bibinfo {pages} {pp.} (\bibinfo {year}
  {2017})}\BibitemShut {NoStop}%
\bibitem [{\citenamefont {Barcelo}\ and\ \citenamefont
  {Visser}(2000)}]{Barcelo:2000zf}%
  \BibitemOpen
  \bibfield  {author} {\bibinfo {author} {\bibfnamefont {C.}~\bibnamefont
  {Barcelo}}\ and\ \bibinfo {author} {\bibfnamefont {M.}~\bibnamefont
  {Visser}},\ }\href {\doibase 10.1088/0264-9381/17/18/318} {\bibfield
  {journal} {\bibinfo  {journal} {Class. Quant. Grav.}\ }\textbf {\bibinfo
  {volume} {17}},\ \bibinfo {pages} {3843} (\bibinfo {year} {2000})},\ \Eprint
  {http://arxiv.org/abs/gr-qc/0003025} {arXiv:gr-qc/0003025 [gr-qc]}
  \BibitemShut {NoStop}%
\bibitem [{\citenamefont {Harko}\ \emph {et~al.}(2013)\citenamefont {Harko},
  \citenamefont {Lobo}, \citenamefont {Mak},\ and\ \citenamefont
  {Sushkov}}]{Harko:2013yb}%
  \BibitemOpen
  \bibfield  {author} {\bibinfo {author} {\bibfnamefont {T.}~\bibnamefont
  {Harko}}, \bibinfo {author} {\bibfnamefont {F.~S.~N.}\ \bibnamefont {Lobo}},
  \bibinfo {author} {\bibfnamefont {M.~K.}\ \bibnamefont {Mak}}, \ and\
  \bibinfo {author} {\bibfnamefont {S.~V.}\ \bibnamefont {Sushkov}},\ }\href
  {\doibase 10.1103/PhysRevD.87.067504} {\bibfield  {journal} {\bibinfo
  {journal} {Phys. Rev.}\ }\textbf {\bibinfo {volume} {D87}},\ \bibinfo {pages}
  {067504} (\bibinfo {year} {2013})},\ \Eprint {http://arxiv.org/abs/1301.6878}
  {arXiv:1301.6878 [gr-qc]} \BibitemShut {NoStop}%
\bibitem [{\citenamefont {Lobo}\ and\ \citenamefont
  {Oliveira}(2009)}]{Lobo:2009ip}%
  \BibitemOpen
  \bibfield  {author} {\bibinfo {author} {\bibfnamefont {F.~S.~N.}\
  \bibnamefont {Lobo}}\ and\ \bibinfo {author} {\bibfnamefont {M.~A.}\
  \bibnamefont {Oliveira}},\ }\href {\doibase 10.1103/PhysRevD.80.104012}
  {\bibfield  {journal} {\bibinfo  {journal} {Phys. Rev.}\ }\textbf {\bibinfo
  {volume} {D80}},\ \bibinfo {pages} {104012} (\bibinfo {year} {2009})},\
  \Eprint {http://arxiv.org/abs/0909.5539} {arXiv:0909.5539 [gr-qc]}
  \BibitemShut {NoStop}%
\bibitem [{\citenamefont {Lobo}(2008)}]{Lobo:2008zu}%
  \BibitemOpen
  \bibfield  {author} {\bibinfo {author} {\bibfnamefont {F.~S.~N.}\
  \bibnamefont {Lobo}},\ }\href {\doibase 10.1088/0264-9381/25/17/175006}
  {\bibfield  {journal} {\bibinfo  {journal} {Class. Quant. Grav.}\ }\textbf
  {\bibinfo {volume} {25}},\ \bibinfo {pages} {175006} (\bibinfo {year}
  {2008})},\ \Eprint {http://arxiv.org/abs/0801.4401} {arXiv:0801.4401 [gr-qc]}
  \BibitemShut {NoStop}%
\bibitem [{\citenamefont {Garcia}\ and\ \citenamefont
  {Lobo}(2010)}]{Garcia:2010xb}%
  \BibitemOpen
  \bibfield  {author} {\bibinfo {author} {\bibfnamefont {N.~M.}\ \bibnamefont
  {Garcia}}\ and\ \bibinfo {author} {\bibfnamefont {F.~S.~N.}\ \bibnamefont
  {Lobo}},\ }\href {\doibase 10.1103/PhysRevD.82.104018} {\bibfield  {journal}
  {\bibinfo  {journal} {Phys. Rev.}\ }\textbf {\bibinfo {volume} {D82}},\
  \bibinfo {pages} {104018} (\bibinfo {year} {2010})},\ \Eprint
  {http://arxiv.org/abs/1007.3040} {arXiv:1007.3040 [gr-qc]} \BibitemShut
  {NoStop}%
\bibitem [{\citenamefont {Montelongo~Garcia}\ and\ \citenamefont
  {Lobo}(2011)}]{MontelongoGarcia:2010xd}%
  \BibitemOpen
  \bibfield  {author} {\bibinfo {author} {\bibfnamefont {N.}~\bibnamefont
  {Montelongo~Garcia}}\ and\ \bibinfo {author} {\bibfnamefont {F.~S.~N.}\
  \bibnamefont {Lobo}},\ }\href {\doibase 10.1088/0264-9381/28/8/085018}
  {\bibfield  {journal} {\bibinfo  {journal} {Class. Quant. Grav.}\ }\textbf
  {\bibinfo {volume} {28}},\ \bibinfo {pages} {085018} (\bibinfo {year}
  {2011})},\ \Eprint {http://arxiv.org/abs/1012.2443} {arXiv:1012.2443 [gr-qc]}
  \BibitemShut {NoStop}%
\bibitem [{\citenamefont {Boehmer}\ \emph {et~al.}(2012)\citenamefont
  {Boehmer}, \citenamefont {Harko},\ and\ \citenamefont
  {Lobo}}]{Bohmer:2011si}%
  \BibitemOpen
  \bibfield  {author} {\bibinfo {author} {\bibfnamefont {C.~G.}\ \bibnamefont
  {Boehmer}}, \bibinfo {author} {\bibfnamefont {T.}~\bibnamefont {Harko}}, \
  and\ \bibinfo {author} {\bibfnamefont {F.~S.~N.}\ \bibnamefont {Lobo}},\
  }\href {\doibase 10.1103/PhysRevD.85.044033} {\bibfield  {journal} {\bibinfo
  {journal} {Phys. Rev.}\ }\textbf {\bibinfo {volume} {D85}},\ \bibinfo {pages}
  {044033} (\bibinfo {year} {2012})},\ \Eprint {http://arxiv.org/abs/1110.5756}
  {arXiv:1110.5756 [gr-qc]} \BibitemShut {NoStop}%
\bibitem [{\citenamefont {Mehdizadeh}\ \emph {et~al.}(2015)\citenamefont
  {Mehdizadeh}, \citenamefont {Kord~Zangeneh},\ and\ \citenamefont
  {Lobo}}]{Mehdizadeh:2015jra}%
  \BibitemOpen
  \bibfield  {author} {\bibinfo {author} {\bibfnamefont {M.~R.}\ \bibnamefont
  {Mehdizadeh}}, \bibinfo {author} {\bibfnamefont {M.}~\bibnamefont
  {Kord~Zangeneh}}, \ and\ \bibinfo {author} {\bibfnamefont {F.~S.~N.}\
  \bibnamefont {Lobo}},\ }\href {\doibase 10.1103/PhysRevD.91.084004}
  {\bibfield  {journal} {\bibinfo  {journal} {Phys. Rev.}\ }\textbf {\bibinfo
  {volume} {D91}},\ \bibinfo {pages} {084004} (\bibinfo {year} {2015})},\
  \Eprint {http://arxiv.org/abs/1501.04773} {arXiv:1501.04773 [gr-qc]}
  \BibitemShut {NoStop}%
\bibitem [{\citenamefont {Capozziello}\ \emph {et~al.}(2012)\citenamefont
  {Capozziello}, \citenamefont {Harko}, \citenamefont {Koivisto}, \citenamefont
  {Lobo},\ and\ \citenamefont {Olmo}}]{Capozziello:2012hr}%
  \BibitemOpen
  \bibfield  {author} {\bibinfo {author} {\bibfnamefont {S.}~\bibnamefont
  {Capozziello}}, \bibinfo {author} {\bibfnamefont {T.}~\bibnamefont {Harko}},
  \bibinfo {author} {\bibfnamefont {T.~S.}\ \bibnamefont {Koivisto}}, \bibinfo
  {author} {\bibfnamefont {F.~S.~N.}\ \bibnamefont {Lobo}}, \ and\ \bibinfo
  {author} {\bibfnamefont {G.~J.}\ \bibnamefont {Olmo}},\ }\href {\doibase
  10.1103/PhysRevD.86.127504} {\bibfield  {journal} {\bibinfo  {journal} {Phys.
  Rev.}\ }\textbf {\bibinfo {volume} {D86}},\ \bibinfo {pages} {127504}
  (\bibinfo {year} {2012})},\ \Eprint {http://arxiv.org/abs/1209.5862}
  {arXiv:1209.5862 [gr-qc]} \BibitemShut {NoStop}%
\bibitem [{\citenamefont {Perlmutter}\ \emph {et~al.}(1999)\citenamefont
  {Perlmutter} \emph {et~al.}}]{Perlmutter:1998np}%
  \BibitemOpen
  \bibfield  {author} {\bibinfo {author} {\bibfnamefont {S.}~\bibnamefont
  {Perlmutter}} \emph {et~al.} (\bibinfo {collaboration} {Supernova Cosmology
  Project}),\ }\href {\doibase 10.1086/307221} {\bibfield  {journal} {\bibinfo
  {journal} {Astrophys. J.}\ }\textbf {\bibinfo {volume} {517}},\ \bibinfo
  {pages} {565} (\bibinfo {year} {1999})},\ \Eprint
  {http://arxiv.org/abs/astro-ph/9812133} {arXiv:astro-ph/9812133 [astro-ph]}
  \BibitemShut {NoStop}%
\bibitem [{\citenamefont {Riess}\ \emph {et~al.}(1998)\citenamefont {Riess}
  \emph {et~al.}}]{Riess:1998cb}%
  \BibitemOpen
  \bibfield  {author} {\bibinfo {author} {\bibfnamefont {A.~G.}\ \bibnamefont
  {Riess}} \emph {et~al.} (\bibinfo {collaboration} {Supernova Search Team}),\
  }\href {\doibase 10.1086/300499} {\bibfield  {journal} {\bibinfo  {journal}
  {Astron. J.}\ }\textbf {\bibinfo {volume} {116}},\ \bibinfo {pages} {1009}
  (\bibinfo {year} {1998})},\ \Eprint {http://arxiv.org/abs/astro-ph/9805201}
  {arXiv:astro-ph/9805201 [astro-ph]} \BibitemShut {NoStop}%
\bibitem [{\citenamefont {Guth}(1981)}]{Guth:1980zm}%
  \BibitemOpen
  \bibfield  {author} {\bibinfo {author} {\bibfnamefont {A.~H.}\ \bibnamefont
  {Guth}},\ }\href {\doibase 10.1103/PhysRevD.23.347} {\bibfield  {journal}
  {\bibinfo  {journal} {Phys. Rev.}\ }\textbf {\bibinfo {volume} {D23}},\
  \bibinfo {pages} {347} (\bibinfo {year} {1981})}\BibitemShut {NoStop}%
\bibitem [{\citenamefont {Linde}(1983)}]{Linde:1983gd}%
  \BibitemOpen
  \bibfield  {author} {\bibinfo {author} {\bibfnamefont {A.~D.}\ \bibnamefont
  {Linde}},\ }\href {\doibase 10.1016/0370-2693(83)90837-7} {\bibfield
  {journal} {\bibinfo  {journal} {Phys. Lett.}\ }\textbf {\bibinfo {volume}
  {129B}},\ \bibinfo {pages} {177} (\bibinfo {year} {1983})}\BibitemShut
  {NoStop}%
\bibitem [{\citenamefont {Martin}\ \emph {et~al.}(2014)\citenamefont {Martin},
  \citenamefont {Ringeval},\ and\ \citenamefont {Vennin}}]{Martin:2013tda}%
  \BibitemOpen
  \bibfield  {author} {\bibinfo {author} {\bibfnamefont {J.}~\bibnamefont
  {Martin}}, \bibinfo {author} {\bibfnamefont {C.}~\bibnamefont {Ringeval}}, \
  and\ \bibinfo {author} {\bibfnamefont {V.}~\bibnamefont {Vennin}},\ }\href
  {\doibase 10.1016/j.dark.2014.01.003} {\bibfield  {journal} {\bibinfo
  {journal} {Phys. Dark Univ.}\ }\textbf {\bibinfo {volume} {5-6}},\ \bibinfo
  {pages} {75} (\bibinfo {year} {2014})},\ \Eprint
  {http://arxiv.org/abs/1303.3787} {arXiv:1303.3787 [astro-ph.CO]} \BibitemShut
  {NoStop}%
\bibitem [{\citenamefont {Wetterich}(1995)}]{Wetterich:1994bg}%
  \BibitemOpen
  \bibfield  {author} {\bibinfo {author} {\bibfnamefont {C.}~\bibnamefont
  {Wetterich}},\ }\href@noop {} {\bibfield  {journal} {\bibinfo  {journal}
  {Astron. Astrophys.}\ }\textbf {\bibinfo {volume} {301}},\ \bibinfo {pages}
  {321} (\bibinfo {year} {1995})},\ \Eprint
  {http://arxiv.org/abs/hep-th/9408025} {arXiv:hep-th/9408025 [hep-th]}
  \BibitemShut {NoStop}%
\bibitem [{\citenamefont {Zlatev}\ \emph {et~al.}(1999)\citenamefont {Zlatev},
  \citenamefont {Wang},\ and\ \citenamefont {Steinhardt}}]{Zlatev:1998tr}%
  \BibitemOpen
  \bibfield  {author} {\bibinfo {author} {\bibfnamefont {I.}~\bibnamefont
  {Zlatev}}, \bibinfo {author} {\bibfnamefont {L.-M.}\ \bibnamefont {Wang}}, \
  and\ \bibinfo {author} {\bibfnamefont {P.~J.}\ \bibnamefont {Steinhardt}},\
  }\href {\doibase 10.1103/PhysRevLett.82.896} {\bibfield  {journal} {\bibinfo
  {journal} {Phys. Rev. Lett.}\ }\textbf {\bibinfo {volume} {82}},\ \bibinfo
  {pages} {896} (\bibinfo {year} {1999})},\ \Eprint
  {http://arxiv.org/abs/astro-ph/9807002} {arXiv:astro-ph/9807002 [astro-ph]}
  \BibitemShut {NoStop}%
\bibitem [{\citenamefont {Barros}\ \emph {et~al.}(2018)\citenamefont {Barros},
  \citenamefont {Amendola}, \citenamefont {Barreiro},\ and\ \citenamefont
  {Nunes}}]{Barros:2018efl}%
  \BibitemOpen
  \bibfield  {author} {\bibinfo {author} {\bibfnamefont {B.~J.}\ \bibnamefont
  {Barros}}, \bibinfo {author} {\bibfnamefont {L.}~\bibnamefont {Amendola}},
  \bibinfo {author} {\bibfnamefont {T.}~\bibnamefont {Barreiro}}, \ and\
  \bibinfo {author} {\bibfnamefont {N.~J.}\ \bibnamefont {Nunes}},\ }\href@noop
  {} {\  (\bibinfo {year} {2018})},\ \Eprint {http://arxiv.org/abs/1802.09216}
  {arXiv:1802.09216 [astro-ph.CO]} \BibitemShut {NoStop}%
\bibitem [{\citenamefont {Butcher}(2015)}]{Butcher:2015sea}%
  \BibitemOpen
  \bibfield  {author} {\bibinfo {author} {\bibfnamefont {L.~M.}\ \bibnamefont
  {Butcher}},\ }\href {\doibase 10.1103/PhysRevD.91.124031} {\bibfield
  {journal} {\bibinfo  {journal} {Phys. Rev.}\ }\textbf {\bibinfo {volume}
  {D91}},\ \bibinfo {pages} {124031} (\bibinfo {year} {2015})},\ \Eprint
  {http://arxiv.org/abs/1503.04145} {arXiv:1503.04145 [gr-qc]} \BibitemShut
  {NoStop}%
\bibitem [{\citenamefont {Kleihaus}\ and\ \citenamefont
  {Kunz}(2014)}]{Kleihaus:2014dla}%
  \BibitemOpen
  \bibfield  {author} {\bibinfo {author} {\bibfnamefont {B.}~\bibnamefont
  {Kleihaus}}\ and\ \bibinfo {author} {\bibfnamefont {J.}~\bibnamefont
  {Kunz}},\ }\href@noop {} {\bibfield  {journal} {\bibinfo  {journal} {Phys.
  Rev.}\ }\textbf {\bibinfo {volume} {D90}},\ \bibinfo {pages} {121503}
  (\bibinfo {year} {2014})}\BibitemShut {NoStop}%
\bibitem [{\citenamefont {Kleihaus}\ and\ \citenamefont
  {Kunz}(2017)}]{Kleihaus:2017kai}%
  \BibitemOpen
  \bibfield  {author} {\bibinfo {author} {\bibfnamefont {B.}~\bibnamefont
  {Kleihaus}}\ and\ \bibinfo {author} {\bibfnamefont {J.}~\bibnamefont
  {Kunz}},\ }\href@noop {} {\bibfield  {journal} {\bibinfo  {journal} {Fundam.
  Theor. Phys.}\ }\textbf {\bibinfo {volume} {189}},\ \bibinfo {pages} {35}
  (\bibinfo {year} {2017})}\BibitemShut {NoStop}%
\bibitem [{\citenamefont {Wongjun}(2017)}]{Wongjun:2017spo}%
  \BibitemOpen
  \bibfield  {author} {\bibinfo {author} {\bibfnamefont {P.}~\bibnamefont
  {Wongjun}},\ }\bibfield  {booktitle} {\emph {\bibinfo {booktitle}
  {{Proceedings, IF-YITP GR+HEP+Cosmo International Symposium VI: Phitsanulok,
  Thailand, August 3-5, 2016}}},\ }\href {\doibase
  10.1088/1742-6596/883/1/012002} {\bibfield  {journal} {\bibinfo  {journal}
  {J. Phys. Conf. Ser.}\ }\textbf {\bibinfo {volume} {883}},\ \bibinfo {pages}
  {012002} (\bibinfo {year} {2017})},\ \Eprint
  {http://arxiv.org/abs/1708.05795} {arXiv:1708.05795 [gr-qc]} \BibitemShut
  {NoStop}%
\bibitem [{\citenamefont {Koivisto}\ and\ \citenamefont
  {Nunes}(2010)}]{Koivisto:2009ew}%
  \BibitemOpen
  \bibfield  {author} {\bibinfo {author} {\bibfnamefont {T.~S.}\ \bibnamefont
  {Koivisto}}\ and\ \bibinfo {author} {\bibfnamefont {N.~J.}\ \bibnamefont
  {Nunes}},\ }\href {\doibase 10.1016/j.physletb.2010.01.051} {\bibfield
  {journal} {\bibinfo  {journal} {Phys. Lett.}\ }\textbf {\bibinfo {volume}
  {B685}},\ \bibinfo {pages} {105} (\bibinfo {year} {2010})},\ \Eprint
  {http://arxiv.org/abs/0907.3883} {arXiv:0907.3883 [astro-ph.CO]} \BibitemShut
  {NoStop}%
\bibitem [{\citenamefont {Morais}\ \emph {et~al.}(2017)\citenamefont {Morais},
  \citenamefont {Bouhmadi-López}, \citenamefont {Sravan~Kumar}, \citenamefont
  {Marto},\ and\ \citenamefont {Tavak}}]{Morais:2016bev}%
  \BibitemOpen
  \bibfield  {author} {\bibinfo {author} {\bibfnamefont {J.}~\bibnamefont
  {Morais}}, \bibinfo {author} {\bibfnamefont {M.}~\bibnamefont
  {Bouhmadi-López}}, \bibinfo {author} {\bibfnamefont {K.}~\bibnamefont
  {Sravan~Kumar}}, \bibinfo {author} {\bibfnamefont {J.}~\bibnamefont {Marto}},
  \ and\ \bibinfo {author} {\bibfnamefont {Y.}~\bibnamefont {Tavak}},\ }\href
  {\doibase 10.1016/j.dark.2016.11.002} {\bibfield  {journal} {\bibinfo
  {journal} {Phys. Dark Univ.}\ }\textbf {\bibinfo {volume} {15}},\ \bibinfo
  {pages} {7} (\bibinfo {year} {2017})},\ \Eprint
  {http://arxiv.org/abs/1608.01679} {arXiv:1608.01679 [gr-qc]} \BibitemShut
  {NoStop}%
\bibitem [{\citenamefont {Koivisto}\ and\ \citenamefont
  {Nunes}(2013)}]{Koivisto:2012xm}%
  \BibitemOpen
  \bibfield  {author} {\bibinfo {author} {\bibfnamefont {T.~S.}\ \bibnamefont
  {Koivisto}}\ and\ \bibinfo {author} {\bibfnamefont {N.~J.}\ \bibnamefont
  {Nunes}},\ }\href {\doibase 10.1103/PhysRevD.88.123512} {\bibfield  {journal}
  {\bibinfo  {journal} {Phys. Rev.}\ }\textbf {\bibinfo {volume} {D88}},\
  \bibinfo {pages} {123512} (\bibinfo {year} {2013})},\ \Eprint
  {http://arxiv.org/abs/1212.2541} {arXiv:1212.2541 [astro-ph.CO]} \BibitemShut
  {NoStop}%
\bibitem [{\citenamefont {Koivisto}\ and\ \citenamefont
  {Nunes}(2009)}]{Koivisto:2009fb}%
  \BibitemOpen
  \bibfield  {author} {\bibinfo {author} {\bibfnamefont {T.~S.}\ \bibnamefont
  {Koivisto}}\ and\ \bibinfo {author} {\bibfnamefont {N.~J.}\ \bibnamefont
  {Nunes}},\ }\href {\doibase 10.1103/PhysRevD.80.103509} {\bibfield  {journal}
  {\bibinfo  {journal} {Phys. Rev.}\ }\textbf {\bibinfo {volume} {D80}},\
  \bibinfo {pages} {103509} (\bibinfo {year} {2009})},\ \Eprint
  {http://arxiv.org/abs/0908.0920} {arXiv:0908.0920 [astro-ph.CO]} \BibitemShut
  {NoStop}%
\bibitem [{\citenamefont {Koivisto}\ \emph {et~al.}(2009)\citenamefont
  {Koivisto}, \citenamefont {Mota},\ and\ \citenamefont
  {Pitrou}}]{Koivisto:2009sd}%
  \BibitemOpen
  \bibfield  {author} {\bibinfo {author} {\bibfnamefont {T.~S.}\ \bibnamefont
  {Koivisto}}, \bibinfo {author} {\bibfnamefont {D.~F.}\ \bibnamefont {Mota}},
  \ and\ \bibinfo {author} {\bibfnamefont {C.}~\bibnamefont {Pitrou}},\ }\href
  {\doibase 10.1088/1126-6708/2009/09/092} {\bibfield  {journal} {\bibinfo
  {journal} {JHEP}\ }\textbf {\bibinfo {volume} {09}},\ \bibinfo {pages} {092}
  (\bibinfo {year} {2009})},\ \Eprint {http://arxiv.org/abs/0903.4158}
  {arXiv:0903.4158 [astro-ph.CO]} \BibitemShut {NoStop}%
\bibitem [{\citenamefont {Sravan~Kumar}\ \emph {et~al.}(2016)\citenamefont
  {Sravan~Kumar}, \citenamefont {Mulryne}, \citenamefont {Nunes}, \citenamefont
  {Marto},\ and\ \citenamefont {Vargas~Moniz}}]{Kumar:2016tdn}%
  \BibitemOpen
  \bibfield  {author} {\bibinfo {author} {\bibfnamefont {K.}~\bibnamefont
  {Sravan~Kumar}}, \bibinfo {author} {\bibfnamefont {D.~J.}\ \bibnamefont
  {Mulryne}}, \bibinfo {author} {\bibfnamefont {N.~J.}\ \bibnamefont {Nunes}},
  \bibinfo {author} {\bibfnamefont {J.}~\bibnamefont {Marto}}, \ and\ \bibinfo
  {author} {\bibfnamefont {P.}~\bibnamefont {Vargas~Moniz}},\ }\href {\doibase
  10.1103/PhysRevD.94.103504} {\bibfield  {journal} {\bibinfo  {journal} {Phys.
  Rev.}\ }\textbf {\bibinfo {volume} {D94}},\ \bibinfo {pages} {103504}
  (\bibinfo {year} {2016})},\ \Eprint {http://arxiv.org/abs/1606.07114}
  {arXiv:1606.07114 [astro-ph.CO]} \BibitemShut {NoStop}%
\bibitem [{\citenamefont {De~Felice}\ \emph
  {et~al.}(2012{\natexlab{a}})\citenamefont {De~Felice}, \citenamefont
  {Karwan},\ and\ \citenamefont {Wongjun}}]{DeFelice:2012jt}%
  \BibitemOpen
  \bibfield  {author} {\bibinfo {author} {\bibfnamefont {A.}~\bibnamefont
  {De~Felice}}, \bibinfo {author} {\bibfnamefont {K.}~\bibnamefont {Karwan}}, \
  and\ \bibinfo {author} {\bibfnamefont {P.}~\bibnamefont {Wongjun}},\ }\href
  {\doibase 10.1103/PhysRevD.85.123545} {\bibfield  {journal} {\bibinfo
  {journal} {Phys. Rev.}\ }\textbf {\bibinfo {volume} {D85}},\ \bibinfo {pages}
  {123545} (\bibinfo {year} {2012}{\natexlab{a}})},\ \Eprint
  {http://arxiv.org/abs/1202.0896} {arXiv:1202.0896 [hep-ph]} \BibitemShut
  {NoStop}%
\bibitem [{\citenamefont {De~Felice}\ \emph
  {et~al.}(2012{\natexlab{b}})\citenamefont {De~Felice}, \citenamefont
  {Karwan},\ and\ \citenamefont {Wongjun}}]{DeFelice:2012wy}%
  \BibitemOpen
  \bibfield  {author} {\bibinfo {author} {\bibfnamefont {A.}~\bibnamefont
  {De~Felice}}, \bibinfo {author} {\bibfnamefont {K.}~\bibnamefont {Karwan}}, \
  and\ \bibinfo {author} {\bibfnamefont {P.}~\bibnamefont {Wongjun}},\ }\href
  {\doibase 10.1103/PhysRevD.86.103526} {\bibfield  {journal} {\bibinfo
  {journal} {Phys. Rev.}\ }\textbf {\bibinfo {volume} {D86}},\ \bibinfo {pages}
  {103526} (\bibinfo {year} {2012}{\natexlab{b}})},\ \Eprint
  {http://arxiv.org/abs/1209.5156} {arXiv:1209.5156 [astro-ph.CO]} \BibitemShut
  {NoStop}%
\bibitem [{\citenamefont {Kumar}\ \emph {et~al.}(2014)\citenamefont {Kumar},
  \citenamefont {Marto}, \citenamefont {Nunes},\ and\ \citenamefont
  {Moniz}}]{Kumar:2014oka}%
  \BibitemOpen
  \bibfield  {author} {\bibinfo {author} {\bibfnamefont {K.~S.}\ \bibnamefont
  {Kumar}}, \bibinfo {author} {\bibfnamefont {J.}~\bibnamefont {Marto}},
  \bibinfo {author} {\bibfnamefont {N.~J.}\ \bibnamefont {Nunes}}, \ and\
  \bibinfo {author} {\bibfnamefont {P.~V.}\ \bibnamefont {Moniz}},\ }\href
  {\doibase 10.1088/1475-7516/2014/06/064} {\bibfield  {journal} {\bibinfo
  {journal} {JCAP}\ }\textbf {\bibinfo {volume} {1406}},\ \bibinfo {pages}
  {064} (\bibinfo {year} {2014})},\ \Eprint {http://arxiv.org/abs/1404.0211}
  {arXiv:1404.0211 [gr-qc]} \BibitemShut {NoStop}%
\bibitem [{\citenamefont {Barros}\ and\ \citenamefont
  {Nunes}(2016)}]{Barros:2015evi}%
  \BibitemOpen
  \bibfield  {author} {\bibinfo {author} {\bibfnamefont {B.~J.}\ \bibnamefont
  {Barros}}\ and\ \bibinfo {author} {\bibfnamefont {N.~J.}\ \bibnamefont
  {Nunes}},\ }\href {\doibase 10.1103/PhysRevD.93.043512} {\bibfield  {journal}
  {\bibinfo  {journal} {Phys. Rev.}\ }\textbf {\bibinfo {volume} {D93}},\
  \bibinfo {pages} {043512} (\bibinfo {year} {2016})},\ \Eprint
  {http://arxiv.org/abs/1511.07856} {arXiv:1511.07856 [astro-ph.CO]}
  \BibitemShut {NoStop}%
\bibitem [{\citenamefont {Barreiro}\ \emph {et~al.}(2017)\citenamefont
  {Barreiro}, \citenamefont {Bertello},\ and\ \citenamefont
  {Nunes}}]{Barreiro:2016aln}%
  \BibitemOpen
  \bibfield  {author} {\bibinfo {author} {\bibfnamefont {T.}~\bibnamefont
  {Barreiro}}, \bibinfo {author} {\bibfnamefont {U.}~\bibnamefont {Bertello}},
  \ and\ \bibinfo {author} {\bibfnamefont {N.~J.}\ \bibnamefont {Nunes}},\
  }\href {\doibase 10.1016/j.physletb.2017.08.061} {\bibfield  {journal}
  {\bibinfo  {journal} {Phys. Lett.}\ }\textbf {\bibinfo {volume} {B773}},\
  \bibinfo {pages} {417} (\bibinfo {year} {2017})},\ \Eprint
  {http://arxiv.org/abs/1610.00357} {arXiv:1610.00357 [gr-qc]} \BibitemShut
  {NoStop}%
\bibitem [{\citenamefont {Turok}\ and\ \citenamefont
  {Hawking}(1998)}]{Turok:1998he}%
  \BibitemOpen
  \bibfield  {author} {\bibinfo {author} {\bibfnamefont {N.}~\bibnamefont
  {Turok}}\ and\ \bibinfo {author} {\bibfnamefont {S.~W.}\ \bibnamefont
  {Hawking}},\ }\href {\doibase 10.1016/S0370-2693(98)00651-0} {\bibfield
  {journal} {\bibinfo  {journal} {Phys. Lett.}\ }\textbf {\bibinfo {volume}
  {B432}},\ \bibinfo {pages} {271} (\bibinfo {year} {1998})},\ \Eprint
  {http://arxiv.org/abs/hep-th/9803156} {arXiv:hep-th/9803156 [hep-th]}
  \BibitemShut {NoStop}%
\bibitem [{\citenamefont {Capozziello}\ \emph {et~al.}(2018)\citenamefont
  {Capozziello}, \citenamefont {Nojiri},\ and\ \citenamefont
  {Odintsov}}]{Capozziello:2018wul}%
  \BibitemOpen
  \bibfield  {author} {\bibinfo {author} {\bibfnamefont {S.}~\bibnamefont
  {Capozziello}}, \bibinfo {author} {\bibfnamefont {S.}~\bibnamefont {Nojiri}},
  \ and\ \bibinfo {author} {\bibfnamefont {S.~D.}\ \bibnamefont {Odintsov}},\
  }\href {\doibase 10.1016/j.physletb.2018.03.064} {\bibfield  {journal}
  {\bibinfo  {journal} {Phys. Lett.}\ }\textbf {\bibinfo {volume} {B781}},\
  \bibinfo {pages} {99} (\bibinfo {year} {2018})},\ \Eprint
  {http://arxiv.org/abs/1803.08815} {arXiv:1803.08815 [gr-qc]} \BibitemShut
  {NoStop}%
\bibitem [{\citenamefont {Capozziello}\ \emph {et~al.}(2014)\citenamefont
  {Capozziello}, \citenamefont {Lobo},\ and\ \citenamefont
  {Mimoso}}]{Capozziello:2013vna}%
  \BibitemOpen
  \bibfield  {author} {\bibinfo {author} {\bibfnamefont {S.}~\bibnamefont
  {Capozziello}}, \bibinfo {author} {\bibfnamefont {F.~S.~N.}\ \bibnamefont
  {Lobo}}, \ and\ \bibinfo {author} {\bibfnamefont {J.~P.}\ \bibnamefont
  {Mimoso}},\ }\href {\doibase 10.1016/j.physletb.2014.01.066} {\bibfield
  {journal} {\bibinfo  {journal} {Phys. Lett.}\ }\textbf {\bibinfo {volume}
  {B730}},\ \bibinfo {pages} {280} (\bibinfo {year} {2014})},\ \Eprint
  {http://arxiv.org/abs/1312.0784} {arXiv:1312.0784 [gr-qc]} \BibitemShut
  {NoStop}%
\bibitem [{\citenamefont {Capozziello}\ \emph {et~al.}(2015)\citenamefont
  {Capozziello}, \citenamefont {Lobo},\ and\ \citenamefont
  {Mimoso}}]{Capozziello:2014bqa}%
  \BibitemOpen
  \bibfield  {author} {\bibinfo {author} {\bibfnamefont {S.}~\bibnamefont
  {Capozziello}}, \bibinfo {author} {\bibfnamefont {F.~S.~N.}\ \bibnamefont
  {Lobo}}, \ and\ \bibinfo {author} {\bibfnamefont {J.~P.}\ \bibnamefont
  {Mimoso}},\ }\href {\doibase 10.1103/PhysRevD.91.124019} {\bibfield
  {journal} {\bibinfo  {journal} {Phys. Rev.}\ }\textbf {\bibinfo {volume}
  {D91}},\ \bibinfo {pages} {124019} (\bibinfo {year} {2015})},\ \Eprint
  {http://arxiv.org/abs/1407.7293} {arXiv:1407.7293 [gr-qc]} \BibitemShut
  {NoStop}%
\bibitem [{\citenamefont {Lobo}(2005{\natexlab{b}})}]{Lobo:2005us}%
  \BibitemOpen
  \bibfield  {author} {\bibinfo {author} {\bibfnamefont {F.~S.~N.}\
  \bibnamefont {Lobo}},\ }\href {\doibase 10.1103/PhysRevD.71.084011}
  {\bibfield  {journal} {\bibinfo  {journal} {Phys. Rev.}\ }\textbf {\bibinfo
  {volume} {D71}},\ \bibinfo {pages} {084011} (\bibinfo {year}
  {2005}{\natexlab{b}})},\ \Eprint {http://arxiv.org/abs/gr-qc/0502099}
  {arXiv:gr-qc/0502099 [gr-qc]} \BibitemShut {NoStop}%
\bibitem [{\citenamefont {Lemos}\ \emph {et~al.}(2003)\citenamefont {Lemos},
  \citenamefont {Lobo},\ and\ \citenamefont {Quinet~de
  Oliveira}}]{Lemos:2003jb}%
  \BibitemOpen
  \bibfield  {author} {\bibinfo {author} {\bibfnamefont {J.~P.~S.}\
  \bibnamefont {Lemos}}, \bibinfo {author} {\bibfnamefont {F.~S.~N.}\
  \bibnamefont {Lobo}}, \ and\ \bibinfo {author} {\bibfnamefont
  {S.}~\bibnamefont {Quinet~de Oliveira}},\ }\href {\doibase
  10.1103/PhysRevD.68.064004} {\bibfield  {journal} {\bibinfo  {journal} {Phys.
  Rev.}\ }\textbf {\bibinfo {volume} {D68}},\ \bibinfo {pages} {064004}
  (\bibinfo {year} {2003})},\ \Eprint {http://arxiv.org/abs/gr-qc/0302049}
  {arXiv:gr-qc/0302049 [gr-qc]} \BibitemShut {NoStop}%
\bibitem [{\citenamefont {Kar}\ and\ \citenamefont
  {Sahdev}(1996)}]{Kar:1995ss}%
  \BibitemOpen
  \bibfield  {author} {\bibinfo {author} {\bibfnamefont {S.}~\bibnamefont
  {Kar}}\ and\ \bibinfo {author} {\bibfnamefont {D.}~\bibnamefont {Sahdev}},\
  }\href {\doibase 10.1103/PhysRevD.53.722} {\bibfield  {journal} {\bibinfo
  {journal} {Phys. Rev.}\ }\textbf {\bibinfo {volume} {D53}},\ \bibinfo {pages}
  {722} (\bibinfo {year} {1996})},\ \Eprint
  {http://arxiv.org/abs/gr-qc/9506094} {arXiv:gr-qc/9506094 [gr-qc]}
  \BibitemShut {NoStop}%
\bibitem [{\citenamefont {Arellano}\ and\ \citenamefont
  {Lobo}(2006)}]{Arellano:2006ex}%
  \BibitemOpen
  \bibfield  {author} {\bibinfo {author} {\bibfnamefont {A.~V.~B.}\
  \bibnamefont {Arellano}}\ and\ \bibinfo {author} {\bibfnamefont {F.~S.~N.}\
  \bibnamefont {Lobo}},\ }\href {\doibase 10.1088/0264-9381/23/20/004}
  {\bibfield  {journal} {\bibinfo  {journal} {Class. Quant. Grav.}\ }\textbf
  {\bibinfo {volume} {23}},\ \bibinfo {pages} {5811} (\bibinfo {year}
  {2006})},\ \Eprint {http://arxiv.org/abs/gr-qc/0608003} {arXiv:gr-qc/0608003
  [gr-qc]} \BibitemShut {NoStop}%
\bibitem [{\citenamefont {Lobo}(2007)}]{Lobo:2007zb}%
  \BibitemOpen
  \bibfield  {author} {\bibinfo {author} {\bibfnamefont {F.~S.~N.}\
  \bibnamefont {Lobo}},\ }in\ \href@noop {} {\emph {\bibinfo {booktitle}
  {Classical and Quantum Gravity Research, 1-78, (2008), Nova Sci. Pub. ISBN
  978-1-60456-366-5}}}\ (\bibinfo {year} {2007})\ \Eprint
  {http://arxiv.org/abs/0710.4474} {arXiv:0710.4474 [gr-qc]} \BibitemShut
  {NoStop}%
\bibitem [{\citenamefont {Bronnikov}(2018)}]{Bronnikov:2017sgg}%
  \BibitemOpen
  \bibfield  {author} {\bibinfo {author} {\bibfnamefont {K.~A.}\ \bibnamefont
  {Bronnikov}},\ }\bibfield  {booktitle} {\emph {\bibinfo {booktitle}
  {{Proceedings, 3rd International Conference on Particle Physics and
  Astrophysics (ICPPA 2017): Moscow, Russia, October 2-5, 2017}}},\ }\href
  {\doibase 10.1142/S0218271818410055} {\bibfield  {journal} {\bibinfo
  {journal} {Int. J. Mod. Phys.}\ }\textbf {\bibinfo {volume} {D27}},\ \bibinfo
  {pages} {1841005} (\bibinfo {year} {2018})},\ \Eprint
  {http://arxiv.org/abs/1711.00087} {arXiv:1711.00087 [gr-qc]} \BibitemShut
  {NoStop}%
\end{thebibliography}%

\end{document}